\documentstyle[emulateapj]{article}

\slugcomment{To appear in PASP (2003 Jan).}

\begin{document}

\title{A Search for Core-Collapse Supernova Progenitors In {\sl Hubble Space 
Telescope\/} Images\footnote{Based on observations made with the NASA/ESA {\sl
Hubble Space Telescope}, obtained from the data archive of the Space Telescope
Science Institute, which is operated by the Association of Universities for
Research in Astronomy, Inc., under NASA contract NAS 5-26555.}}

\author{Schuyler D.~Van Dyk}
\affil{IPAC/Caltech, Mailcode 100-22, Pasadena CA  91125}
\authoremail{vandyk@ipac.caltech.edu}

\author{Weidong Li and Alexei V.~Filippenko}
\affil{Department of Astronomy, 601 Campbell Hall, University of
California, Berkeley, CA  94720-3411}
\authoremail{alex@astro.berkeley.edu, weidong@astro.berkeley.edu}

\begin{abstract}
Identifying the massive progenitor stars that give rise to core-collapse
supernovae (SNe) is one of the main pursuits of supernova and stellar evolution
studies.  Using ground-based images of recent, nearby SNe obtained primarily
with the Katzman Automatic Imaging Telescope, astrometry from the Two Micron
All Sky Survey, and archival images from the {\sl Hubble Space Telescope}, we
have attempted the direct identification of the progenitors of 16 Type II and
Type Ib/c SNe.  We may have identified the progenitors of the Type II SNe
1999br in NGC 4900, 1999ev in NGC 4274, and 2001du in NGC 1365 as supergiant
stars with $M^0_V\approx -6$ mag in all three cases.  We may have also
identified the progenitors of the Type Ib SNe 2001B in IC 391 and 2001is in NGC
1961 as very luminous supergiants with $M^0_V \approx -8$ to $-9$ mag, and
possibly the progenitor of the Type Ic SN 1999bu in NGC 3786 as a supergiant
with $M^0_V\approx -7.5$ mag.  Additionally, we have recovered at late times
SNe 1999dn in NGC 7714, 2000C in NGC 2415, and 2000ew in NGC 3810, although
none of these had detectable progenitors on pre-supernova images.  In fact, for
the remaining SNe only limits can be placed on the absolute magnitude and color
(when available) of the progenitor.  The detected Type II progenitors and
limits are consistent with red supergiants as progenitor stars, although
possibly not as red as we had expected.  Our results for the Type Ib/c SNe do
not strongly constrain either Wolf-Rayet stars or massive interacting binary
systems as progenitors.
\end{abstract}


\section{Introduction}

Determining the progenitor stars that give rise to supernovae (SNe) is at the
heart of SN research and is certainly a key aspect of stellar evolution
studies.  Without knowledge of the nature of SN progenitors, many of the
conclusions and inferences that have been made from SNe on the chemical
evolution of galaxies, the energy input into the interstellar medium, the
production of stellar remnants such as neutron stars and black holes, the
origin of cosmic rays, and even the determination of cosmological distances
stand on precarious ground.  The main obstacle is that a SN leaves few traces
of the star that exploded.  Additionally, only a small handful of the more than
2000 historical SNe have had pre-explosion objects identified.  These include
SN 1961V in NGC 1058 (Zwicky 1964, 1965), SN 1978K in NGC 1313 (Ryder et
al.~1993), SN 1987A in the LMC (e.g., Gilmozzi et al.~1987; Sonneborn, Altner,
\& Kirshner 1987), SN 1993J in M81 (Aldering, Humphreys, \& Richmond 1994;
Cohen, Darling, \& Porter 1995), and SN 1997bs in M66 (Van Dyk et al. 1999b,
2000).  It should be noted that these five SNe were all at least somewhat
unusual, and both SN 1961V (Goodrich et al.~1989; Filippenko et al.~1995; but
see also Van Dyk, Filippenko, \& Li 2002a) and SN 1997bs (Van Dyk et al. 2000)
may not have been actual SNe (defined to be the catastrophic explosion of a
star at the end of its life).

SNe are characterized optically by the presence or absence of H in their
spectra near maximum brightness: the Type II SNe (SNe~II) and Type I SNe
(SNe~I), respectively.  SNe~I further divide into SNe~Ia, which are
characterized by a deep absorption trough around 6150~\AA\ produced by
blueshifted Si~II $\lambda$6355, and SNe~Ib/c, which do not show this trough.
SNe~Ib exhibit strong He~I absorption, while SNe~Ic show little or no evidence
for He~I absorption.  The SNe~II also include various subtypes: SNe II-plateau
(II-P) and II-linear (II-L), based on the shape of their light curves (but with
associated spectral characteristics as well; Schlegel 1996; Filippenko 1997),
and SNe II-narrow (IIn), which lack the blueshifted component of the broad
Balmer-line P-Cygni profiles, but instead show a relatively narrow emission
component atop the broad one.  SNe IIb may be a bridge between the SNe II and
Ib/c, possessing properties of each. (See Filippenko 1997 for a thorough review
of SN spectra and types.)

SNe Ia are thought to arise from the thermonuclear deflagration and/or
detonation of a white dwarf, but they are not the focus of this paper.  SNe II
and Ib/c probably arise from the collapse of the Fe core toward the end of the
life of a massive ($M{\gtrsim}10\ M_{\odot}$) star.  Whereas it is generally
agreed that SNe II must arise from the explosions of hydrogen-rich supergiant
stars, the progenitors of SNe Ib/c have not been unambiguously identified.
Clearly, SNe Ib/c must arise from stars that have lost most or all of their
hydrogen envelopes.  As such, Wolf-Rayet stars have been proposed as possible
progenitors (see Branch, Nomoto, \& Filippenko 1991, and references therein).
Alternatively, Uomoto (1986), Nomoto, Filippenko, \& Shigeyama (1990),
Podsiadlowski, Joss, \& Hsu (1992), Iwamoto et al.~(1994), and Nomoto et
al.~(1996) have explored massive gas-transferring binary systems as possible SN
Ib/c progenitors.  (Another possibility, now considered much less likely, are
off-center explosions of white dwarfs; Branch \& Nomoto 1986.)

Part of the evidence for the core-collapse nature of SNe~II and Ib/c comes from
theoretical modelling (e.g., Woosley \& Weaver 1986, 1995), but indications
that these SNe have massive-star progenitors also stems from the few that have
had progenitors directly identified: for SN 1961V, an extremely luminous
($M_{\rm pg}^0 \approx -12$ mag) star (Bertola 1964; Zwicky 1964; Klemola
1986); for SN IIn 1978K, a reddish star, with $B-R \approx 2$ mag and $M_B
\approx -6$ mag (Ryder et al.~1993); for SN II-P 1987A, a massive blue
supergiant (e.g., Woosley 1988); for SN IIb 1993J, a red supergiant, possibly
in a binary system (Podsiadlowski et al.~1993; Aldering et al.~1994; Van Dyk et
al.~2002b); and for SN IIn 1997bs, a supergiant star with $M_V\approx -7.4$ mag
(Van Dyk et al.~1999b, 2000).  In addition, young SN remnants, such as Cas A
(e.g., Fesen, Becker, \& Blair 1987; Fesen \& Becker 1991; Garc{\'\i}a-Segura,
Langer, \& Mac Low 1996), clearly point toward massive progenitors.

Evidence for massive progenitors has also been accrued from the environmental
data for many SNe.  Van Dyk (1992) and Van Dyk, Hamuy, \& Filippenko (1996)
provided statistics of the association of SNe II and Ib/c with massive
star-formation regions, from ground-based imaging.  More recently, Barth et
al.~(1996) and Van Dyk et al.~(1999a,b) have exploited the superior spatial
resolution afforded by the {\sl Hubble Space Telescope\/} ({\sl HST}) to
resolve individual stars in SN environments and place constraints on the
progenitor ages and masses.  Based on the properties of the surrounding stellar
association, Van Dyk et al.~(1999a), in particular, concluded that the
progenitor of the SN II-L 1979C in M100 had an initial mass $M\approx 17$--18
$M_{\odot}$.  Using {\sl HST\/} images of SN 1993J in M81 to remove
contamination by neighboring stars of the ground-based estimates of the
progenitor brightness (Aldering et al.~1994), Van Dyk et al.~(2002b) constrain
the progenitor mass to be $\sim$13--22 $M_{\odot}$.

Clearly, direct identification of the progenitors of additional core-collapse
SNe is essential.  Van Dyk et al.~(1999b, 2000) were able to directly identify
the progenitor star for SN 1997bs using {\sl HST\/} archival images.  However,
at that point in time the quantity of archival data in which pre-SN images
might exist for recent SNe was extremely small.  We can now reap the benefits
from the confluence of two circumstances: the increasing data volume in the
{\sl HST\/} archive and the success of modern SN search programs, in
particular, the Lick Observatory SN Search (LOSS; Li et al. 1999; Filippenko et
al. 2001) and the Lick Observatory and Tenagra Observatory SN Searches (LOTOSS;
Schwartz et al. 2000; Beutler et al. 2002), with which two of us (W.D.L. \&
A.V.F.) are involved and which are discovering new SNe at a remarkable rate
(e.g., 65 in 2002 January--September).

In this paper we discuss our attempt to isolate the progenitors of 16
core-collapse SNe (6 SNe~II and 10 SNe~Ib/c) using {\sl HST\/} Wide Field
Planetary Camera 2 (WFPC2) images of galaxies.  It will only be through the
accumulation of a statistically significant number of direct identifications of
progenitor stars for both SNe~Ib/c and SNe~II that we finally will be able to
adequately test the various models for massive stellar evolution and inevitable
explosion.

\section{Method of Analysis}

We began by cross-referencing historical SNe with the {\sl HST\/} archive.  We
compiled a list of core-collapse events, all since about 1997 through 2002
June, which might contain the progenitor star in at least one WFPC2 image.  A
summary of the available data is in Table 1.  The crux of this work is
determining at which location on the four chips of the image array the star
should be.  It is therefore of utmost importance to have high astrometric
accuracy for all the images.  Ideally, one could pinpoint the exact SN location
by comparing a late-time image of the SN with a pre-SN image.  In fact, in
three cases below, we have been able to do this.  However, even locating the
fading SN in {\sl HST\/} images is often quite difficult and requires high
astrometric precision.

It is reasonably straightforward to measure accurate positions (to fractions of
an arcsec) for SNe on high-quality ground-based images.  However, it is well
known that positions based on the astrometric information in the {\sl HST\/}
image headers alone are not very accurate.  Online documentation\footnote{
http://www.stsci.edu/instruments/wfpc2/Wfpc2{\textunderscore}faq/wfpc2{\textunderscore}ast{\textunderscore}faq.html.}
for WFPC2 claims an accuracy of ${\sim}0{\farcs}5$; from experience, however,
we have found it to be more typically ${\sim}1{\farcs}5$ {\it or worse} (see,
e.g., Filippenko et al.~1995; Van Dyk et al.~1999b).  It is thus incumbent upon
us to apply an {\it independent\/} astrometric grid to the WFPC2 images.

For this reason we have adopted the Two Micron All Sky Survey (2MASS) as the
basis for the astrometric grid for both the ground-based SN images and the {\sl
HST\/} images potentially containing the SN progenitor. While it is well known
that the 2MASS near-infrared catalogs are an unprecedented photometric
resource, a less recognized fact is that the Point Source Catalog is also of
outstanding astrometric quality, with residuals in the final Catalog typically
and conservatively $0{\farcs}10$ (H. McCallon \& R. Cutri 2002, private
communication).  The $J$-band 2MASS images correspond quite well with the
optical SN and {\sl HST\/} images, such that a sufficient number of astrometric
fiducial stars on the optical images can be employed to achieve accuracies in
the astrometric solutions of typically $0{\farcs}2$--$0{\farcs}3$.  For WFPC2
this means, with a plate scale of $0{\farcs}1$ pixel$^{-1}$, that 
we can estimate the position of the progenitor star on a WF chip 
with uncertainty in the range of only three to nine pixels (roughly double this 
pixel range for the PC chip).

Unless otherwise specified, we have measured all of the SN positions from
$R$-band or unfiltered images obtained by the Katzman Automatic Imaging
Telescope (KAIT; Filippenko et al.~2001; KAIT is the principal instrument for
LOSS/LOTOSS), as part of KAIT SN light-curve monitoring (see, e.g., Li et
al.~2001; Modjaz et al.~2001b; Leonard et al.~2002a,b).  We then use the STSDAS
routine {\it wmosaic\/} within IRAF{\footnote{IRAF (Image Reduction and
Analysis Facility) is distributed by the National Optical Astronomy
Observatories, which are operated by the Association of Universities for
Research in Astronomy, Inc., under cooperative agreement with the National
Science Foundation.}} to stitch together the four WFPC2 chips to obtain the
full WFPC2 field of view.  (This step is generally necessary for there to be
enough available fiducial stars to facilitate the astrometry.)  The {\it
wmosaic\/} routine makes corrections for geometric distortion in each chip and
for the rotation, offsets, and scale differences among the chips.  Application
of 2MASS to the full mosaic results in an astrometric reference frame, or grid,
of satisfactorily high accuracy, verified by the resulting positions of at
least one or more ``check stars'' on the mosaic. (The key, again, is that we
are not employing the {\sl HST\/} image header information, but imposing our
own astrometric grid on the mosaic.)

Often we could match unsaturated stars on the WFPC2 mosaic with 2MASS sources
and directly derive an astrometric solution for the mosaic.  When this was not
possible, owing to a lack of 2MASS sources seen on the WFPC2 mosaic, we matched
2MASS sources first with relatively bright (but unsaturated) stars on deep
optical images (typically $V$ band) of the SN host galaxies (obtained in 2002
February and April at the Palomar 1.5-m telescope) and then, once an
astrometric solution had been obtained, matched fainter stars seen in both the
deep ground-based images and the WFPC2 mosaic, obtaining a new astrometric
solution based on these objects.  All astrometric solutions were derived
throughout using the IRAF task {\it ccmap\/} and applied using {\it cctran}.
Finally, we then determined the individual WFPC2 chip pixel value and the
uncertainty in that value for the SN site and, therefore, for the progenitor.

Once the site was located, photometry of the appropriate WFPC2 chip was
performed using the routine HSTphot (Dolphin 2000a,b), which automatically
accounts for WFPC2 point-spread function (PSF) variations and charge-transfer
effects across the chips, zeropoints, aperture corrections, etc., and can
return magnitudes in standard Johnson-Cousins bands as output, whenever
possible.  HSTphot was run in all cases with a 3$\sigma$ detection threshold.
Dolphin (2000a) tested HSTphot against DoPHOT on the same dataset and found no
systematic differences in the results from the two packages.  Similarly, Saha
et al.~(2001), in their measurement of the Cepheid distance to NGC 3982, find
the HSTphot results to be within the errors of the DoPHOT results.  Extensive
tests by D.~C.~Leonard (2002, private communication) of DAOPHOT against HSTphot
also show very good agreement (in particular, on images of NGC 3351, he finds
${\delta}V = 0.016$ and ${\delta}I = 0.043$ mag; see Graham et al.~1997).
Consequently, we can determine the magnitude and, when available, the color of
the candidate progenitor, although, as we will see, for most objects this turns
out to be placing only a {\it limit} on the magnitude and color.

In three cases we were able to recover the SN in a late-time image, allowing us
to isolate the exact position of the progenitor.  These detections serve as a
valuable test of our method: if the astrometric grid is truly accurate, we
should be able to locate the old SN in a straightforward manner, which, in
fact, proves to be the case.  We emphasize that, short of having an accurate
absolute position for these SNe, the generally faint sources would be
relatively difficult to unambiguously locate on a WFPC2 chip, especially with
the possible presence of other variable stars in the host galaxy.  The
coincidence of the estimated SN position on the chip with the actual recovery
position provides us with some confidence in the progenitor positions for the
other cases, where no late-time image of the SN is available.

All quoted positions are J2000 throughout.  All observing dates are in UT, and
the relevant {\sl HST\/} GO or GTO program is given in each case.  Unless
otherwise specified, the distance to the host galaxy is derived from the
heliocentric radial velocity corrected for Local Group infall into the Virgo
Cluster given in the Lyon-Meudon Extragalactic Database (LEDA), and is based on
an assumed distance scale of 65 km s$^{-1}$ Mpc$^{-1}$.  For lack of other
data, the extinction correction toward the SN is limited in most cases to the
Galactic component [generally adopting $A_V$ from the NASA/IPAC Extragalactic
Database (NED), and originally from Schlegel, Finkbinder, \& Davis (1998)],
assuming the Cardelli, Clayton, \& Mathis (1989) reddening law.  All limiting
magnitudes and colors, unless otherwise specified, are based on the 3$\sigma$
detection limits.  Below we discuss the individual core-collapse SNe and their
possible progenitors.

\section{The Supernovae and Their Progenitors}

\subsection{SN 1999an in IC 755}

SN 1999an was discovered by Wei et al.~(1999) as part of the Beijing
Astronomical Observatory (BAO) SN Search and was classified as a SN II by Cao
\& Gu (1999).  We measure the SN position from a KAIT image as $\alpha$ =
12$^h$01$^m$10${\fs}$57, $\delta$ = +14{\arcdeg}06{\arcmin}11${\farcs}$1, with
uncertainty ${\pm}0{\farcs}4$.  The host galaxy was imaged in F606W (160 s) by
GO-5446 on 1995 January 5.  The Galactic foreground stars around the host on
the WFPC2 mosaic are quite faint, and so we used a deep Palomar $V$-band image
to establish the astrometric grid for the mosaic.  The uncertainty in the
positions for these stars in the Palomar image is ${\pm}0{\farcs}2$.  However,
applying the grid to the mosaic resulted in an uncertainty of
${\pm}0{\farcs}6$, likely due to the relative faintness of the stars in the
mosaic. Together with the measured uncertainty in the SN position, this results
in a total uncertainty of ${\pm}0{\farcs}8$ in the SN position on the mosaic
(the uncertainties in the measured SN position and in the astrometric grid were
added in quadrature here and throughout).

Figure 1 shows the SN site on the WF2 chip.  Within the error circle are six
objects, with $m_{\rm F606W}\approx 21.9$--22.9 mag.  However, HSTphot
considers them all to be spatially extended (non-stellar).  Assuming a host
galaxy distance of 23 Mpc (based on Tully-Fisher estimates by Yasuda, Fukugita,
\& Okamura 1997) and Galactic $A_V=0.1$ mag, these objects have $M^0_V\approx
-9.0$ to $-10.0$ mag, which are at the extreme end of observed stellar
luminosity in the Milky Way (e.g., Humphreys \& Davidson 1979).  It is
therefore likely that these are compact star clusters.  The progenitor may have
been a member of one of these clusters, and we are unable to resolve it.  It is
also possible that the progenitor was not a member and also not detected in the
image.  If the latter is the case, then from the detection limit in the SN
environment ($m_{\rm F606W}\gtrsim 25.0$ mag) we can place an upper limit on
the progenitor's absolute visual magnitude of $M^0_V \gtrsim -6.9$ mag.

\subsection{SN 1999br in NGC 4900}

SN 1999br was discovered by King (1999) as part of LOSS and was classified as a
SN II by Garnavich et al.~(1999a) and Filippenko, Stern, \& Reuland (1999); the
latter point out that the SN is very subluminous, regardless of its type.
Patat et al.~(1999) and Zampieri et al.~(2002) conclude that a smaller than
``normal'' $^{56}$Ni mass must have been produced in the explosion.
Nonetheless, Hamuy \& Pinto (2002) consider SN 1999br as a SN II-P.  We measure
the SN position from a KAIT image as $\alpha$ = 13$^h$00$^m$41${\fs}$82,
$\delta$ = +02{\arcdeg}29{\arcmin}45${\farcs}$4, with uncertainty
${\pm}0{\farcs}2$.  The host galaxy was imaged in F606W (160 s) by GO-5446 on
1995 January 29.  The stars useful for astrometry in the WFPC2 mosaic are all
quite faint (the bright star near the SN position is saturated); therefore, we
used a deep Palomar $V$-band image to establish the astrometric grid for the
mosaic, with uncertainty ${\pm}0{\farcs}3$, which results in a total
uncertainty of ${\pm}0{\farcs}5$.

Figure 2 shows the SN site on the WF3 chip.  A very faint object is seen within
the error circle, but it is not detected by HSTphot.  A brighter object is
detected along the north-northeast edge of the circle, with $m_{\rm
F606W}=24.96 {\pm} 0.27$ mag.  Tentatively, we assign this as the possible
progenitor of SN 1999br (it is detected in the F606W image at the 4$\sigma$
level).  The host galaxy has also been very recently (on 2002 June 20) and more
deeply imaged by GO-9042 in F450W and F814W (460 s each).  It is therefore
possible that we can detect the SN at late times and pinpoint the exact
progenitor location.  We applied the same grid as on the F606W mosaic to the
F450W and F814W mosaics, using four of the same five faint fiducial stars.

Figure 3 illustrates the SN site on the PC chip and also shows that the
astrometry produces quite similar results as in Figure 2 for the SN location.
However, no object is detected within the error circle to $B \gtrsim 25.3$ and
$I \gtrsim 24.5$ mag.  This is consistent with an extrapolation from the
late-time SN photometry (SN 1999br was last detected in 2000 April at $B\approx
23.6$ and $I\approx 21.2$ mag, and, following the $^{56}$Co decay rate, a
decline of $\sim$8 mag is expected between 2000 April and 2002 June;
A.~Pastorello 2002, private communication).  We therefore most likely have {\it
not\/} recovered the SN in these 2002 images. However, the fact that the
possible progenitor object identified above is no longer detected in the deep
F450W and F814W images indicates that we may have identified the SN progenitor
in the older (1995) F606W image.

For a distance of 17.3 Mpc (Ho, Filippenko, \& Sargent 1997) and Galactic
extinction toward the host galaxy $A_V=0.08$ mag [Hamuy \& Pinto (2002) list
$A_V \approx 0$ mag within the host, based on the SN color at the end of the
plateau], we find that $M^0_V \approx -6.3$ mag for this star, consistent with
that for late-type (M-type) supergiants (Humphreys \& Davidson 1979).
Unfortunately, no color information exists for this star.  If it is not the
progenitor, we can place a limit on the progenitor's absolute magnitude of
$M^0_V \gtrsim -5.9$ mag, based on the F606W image detection threshold ($m_{\rm
F606W}\gtrsim 25.4$ mag).

\subsection{SN 1999bu in NGC 3786}

SN 1999bu was discovered by Li (1999) using KAIT, about 1\arcsec\ W and
3\arcsec\ S of the host galaxy nucleus.  Jha et al.~(1999a) classified it as a
SN Ic.  NGC 3786 was imaged by GO-5479 in a single 500-s F606W exposure on 1995
March 30.  Only two stars are seen in the WFPC2 mosaic.  For this reason, we
performed an offset from a relatively bright (but unsaturated) star
$11{\farcs}96$ W and $11{\farcs}90$ N of the SN site.  This offset was
determined from a KAIT image of the SN, after galaxy background light had been
subtracted to make it easier to derive a centroid for the SN.  We applied the
same offset from this star on the WFPC2 mosaic, and we show the SN position on
the PC chip in Figure 4.  We have adopted an uncertainty in the offset which is
the uncertainty in the astrometry of the KAIT image, namely ${\pm}0{\farcs}5$.

Three objects are detected within and along the northern edge of the error
circle.  Object A is considered extended by HSTphot and has $m_{\rm
F606W}=23.62{\pm}0.01$ mag. Object B is also considered extended and has $m_{\rm
F606W}=23.54{\pm}0.01$ mag. Object C, just east of north along the circle edge, is
considered stellar and has $m_{\rm F606W}=25.72{\pm}0.31$ mag.

For a distance of about 42 Mpc [correcting the distance of 36.1 Mpc from Pogge
\& Martini (2002) to our assumed distance scale] and Galactic extinction
$A_V=0.08$ mag, objects A and B, presumably small star clusters, have
$M^0_V\approx -9.6$ and $-9.7$ mag, respectively, brighter than known
supergiants.  Object C, however, has $M^0_V\approx -7.5$ mag, consistent with
supergiant brightnesses; although it is at the edge of the error circle, it
could possibly be the progenitor.  Alternatively, the progenitor may have been
a member of the two clusters and is unresolved.  It is also possible that it
was not object C and was not a cluster member, and was not detected.  If the
latter is the case, then from the detection limit in the SN environment
($m_{\rm F606W}\gtrsim 26.1$ mag), we estimate $M^0_V \gtrsim -7.1$ mag for the
progenitor.

\subsection{SN 1999bx in NGC 6745}

SN 1999bx was discovered using KAIT by Friedman \& Li (1999).  The host galaxy,
NGC 6745 (UGC 11391), may actually be an interacting galaxy pair, and the SN
occurred about 2\arcsec\ W and 15\arcsec\ N of the southern of the two nuclei.
Jha et al.~(1999b) classified it as a SN~II.  From a KAIT image, we measure the
SN position as $\alpha$ = 19$^h$01$^m$41${\fs}$39, $\delta$ =
+40{\arcdeg}44{\arcmin}52${\farcs}$0, with uncertainty ${\pm}0{\farcs}2$.  NGC
6745 was imaged very deeply in F336W (22000 s), F555W (4800 s), F675W (5200 s),
and F814W (5200 s) on 1997 March 19 and 21 by GO-6276.  Applying the 2MASS
astrometric grid directly to the F555W mosaic, with total positional
uncertainty ${\pm}0{\farcs}4$, we show in Figure 5 the SN site on the WF3 chip.

We detect four objects (A--D) within, or along the edge of, the error circle,
and with the good color coverage discussed above, we can derive useful
information about them.  HSTphot considers the two objects at the western
circle edge, A and B, to be stellar, with $V=24.37{\pm}0.09$ and
$25.01{\pm}0.20$ mag, respectively, whereas the two eastern objects, C and D,
are likely extended, with $V=23.75{\pm}0.01$ and $23.53{\pm}0.01$ mag,
respectively.  Assuming a distance of about 74 Mpc and Galactic extinction
$A_V=0.44$ mag, the two western objects, A and B, have $M_V=-10.4$ and $-9.8$
mag, respectively, certainly at the upper end of possible stellar luminosities,
while the two eastern objects, C and D, have $M_V=-11.0$ and $-11.3$ mag,
respectively, almost certainly too bright to be single stars.  Additionally,
Objects A and B are quite blue: for A, the reddening-corrected colors are
$(U-V)^0=-0.38$, $(V-R)^0=0.01$, $(R-I)^0=0.19$, and $(V-I)^0=0.20$ mag; and,
for Object B, $(U-V)^0= -0.56$, $(V-R)^0=0.60$, $(R-I)^0=0.20$, and
$(V-I)^0=0.80$ mag.  Colors for Object B indicate that it may be a composite of
blue and red objects.

It is therefore likely that the detected objects are too luminous and too blue
to be the possible progenitor of SN 1999bx, since as a SN II, we might expect
the progenitor to have been a red supergiant. We cannot rule out that the
progenitor was either an unresolved member of likely clusters C and D, or was
blended with luminous stellar Objects A or B. However, if it is undetected,
limits on the absolute magnitude and color of the progenitor are $M^0_V \gtrsim
-7.8$, $(U-V)^0 \lesssim 0.5$, $(V-R)^0 \lesssim 0.9$, $(R-I)^0 \lesssim 1.0$,
and $(V-I)^0 \lesssim 1.6$ mag.  (Color limits here and throughout are derived
from the larger SN environment, $\sim$10\arcsec\ or so, depending on the image,
and are not particularly restrictive.)

\subsection{SN 1999dn in NGC 7714}

SN 1999dn was discovered by Qiu et al.~(1999) as part of the BAO SN Search. It
was classified as a SN Ic by Ayani et al.~(1999) and Turatto et al.~(1999).
Pastorello et al.~(1999) describe the strong resemblance of SN 1999dn to SNe
1997X, 1994I, and 1996aq around maximum brightness, with He~I lines detected;
they argue that it should be considered as a SN~Ib/c.  Deng et al.~(2000) and
Matheson et al.~(2001a) further refine the classification to Type Ib; Branch et
al.~(2002) consider it the currently best-observed, ``fiducial'' SN Ib.  Deng
et al. (2000) also find strong evidence for both H$\alpha$ and C II
$\lambda$6580.  They conclude that the low-mass H skin above the He layer in SN
1999dn makes this event a possible link between SNe Ib and IIb, such as SN
1993J (e.g., Filippenko, Matheson, \& Ho 1993).  From a KAIT image we measure
the SN position as $\alpha$ = 23$^h$36$^m$14${\fs}$81, $\delta$ =
+02{\arcdeg}09{\arcmin}08${\farcs}$4, with uncertainty ${\pm}0{\farcs}5$.

SN 1999dn is potentially one of the more interesting objects in this study,
since we can attempt to detect the faint SN in one of our own {\sl HST\/}
Snapshot images (GO-8602; see Li et al.~2002, which does not include SN 1999dn
in the analysis) and compare this with the pre-SN archive image.  We obtained a
700-s F814W image (cosmic-ray split pair) on 2001 January 24, and 700-s splits
each in F555W and F814W on 2001 July 10.  However, this turned out to be one of
the most difficult sets of {\sl HST\/} images for which to establish an
astrometric grid, due to the lack of stars in common between these images and
2MASS detections (unfortunately, we did not observe the host galaxy at
Palomar).  Nonetheless, we can apply a grid, with ${\pm}0{\farcs}4$
uncertainty, and locate the SN on the PC chip for the first pair of F814W
exposures from 2001 January, with total uncertainty ${\pm}0{\farcs}6$.  The SN
is most likely the faint object just east of the error circle center in
Figure 6 (we have applied the routine {\it qzap\/}, written by M.~Dickinson, 
to the image in the figure, to remove residual cosmic-ray hits; the
object to the south is seen at about the same brightness in the second F814W
image pair from 2001 July).  The SN had $m_{\rm F814W} = 24.18{\pm}0.18$ mag.
It is undetected in either band of the second set of Snapshot images, to $V
\gtrsim 26.2$, $V-I \lesssim 2.0$ mag.  (Two 300-s F300W images were also
obtained by GO-9124 on 2001 August 3, but the SN had most likely faded well
below detection in these images, and so we do not further consider them.)

The site is also in a single 500-s F606W exposure obtained by GO-5479 on 1996
May 15 and in four F380W images (total exposure time 1800~s) obtained by
GO-6672 on 1998 August 29.  We show in Figure 7 the SN site in the F606W image
(on the PC chip; again, we have applied {\it qzap\/} to make the image more 
cosmetically appealing in the figure).
Nothing is detected at the SN site to $m_{\rm F606W} \gtrsim 26.0$ mag.
Additionally, nothing is detected in the F380W image to $m_{\rm F380W} \gtrsim
24.7$ mag.  For a distance of about 43 Mpc and Galactic $A_V=0.17$ mag [Turatto
et al. (1999) indicate no extinction to the SN], the corrected magnitude and
color correspond to $M^0_V \gtrsim -7.3$, $(U-V)^0 \lesssim 2.5$ mag for the
progenitor.

\subsection{SN 1999ec in NGC 2207}

SN 1999ec was discovered using KAIT by Modjaz \& Li (1999). The SN was
classified as a SN Ib by Jha et al.~(1999c), although Matheson et al.~(2001a)
consider it a peculiar SN~I, without a well-defined type, similar to SN~1993R
(cf.~Filippenko \& Matheson 1993).  We measure on a KAIT image the SN position as
$\alpha$ = 06$^h$16$^m$16${\fs}$18, $\delta$ =
$-$21{\arcdeg}22{\arcmin}10${\farcs}$1, with uncertainty ${\pm}0{\farcs}2$.
[This position is discrepant by at least 1\arcsec\ with that referred to by
Elmegreen et al. (2001).]  The host galaxy is part of an interacting system,
extensively studied by Elmegreen et al.~with {\sl HST\/} (GO-6483); only their
``NGC2207-NW'' F336W (2000 s), F439W (2000 s), F555W (660 s), and F814W (720 s)
images from 1996 May 25 are of use here.  We apply the 2MASS astrometric grid
directly to the WFPC2 F555W mosaic, with uncertainty ${\pm}0{\farcs}2$.  Figure
8 shows the SN site on the F555W WF2 chip, with total uncertainty
${\pm}0{\farcs}3$.

A hint of an object is possibly seen within the circle, but nothing is detected
there by HSTphot to $V \gtrsim 25.9$ mag.  The Galactic extinction is
$A_V=0.29$ mag.  To derive possibly more accurate limits on the brightness and
color of a progenitor star, we exploit the color information available to us
for objects in the SN's environment, to possibly estimate the local extinction.
As can be seen from the color-color diagram in Figure 9, HSTphot considers most
of the objects around the error circle to be extended; they are likely star
clusters.  However, two objects are considered stellar, and the one (Star 1)
closest to the SN position implies that $A_V \approx 1.6$ mag (though the
$U$-band photometry may be the least certain, the extinction, even for the
clusters, appears to range from $\sim$1 to $\sim$3 mag).  Although it is at the
edge of the error circle, it is possible that Star 1, with $V=23.42$,
$U-B=-0.92$, $B-V=0.43$, and $V-I=0.38$ mag, is the progenitor.  Adjusting the
distance to the host from Elmegreen et al.~(2001) to about 40 Mpc for our
adopted distance scale, and assuming the $A_V$ derived from Star 1, this leads
to the star having a very blue intrinsic color [$(U-B)^0=-1.3$, $(B-V)^0=-0.1$,
and $(V-I)^0=-0.2$ mag], but also $M^0_V\approx -11.2$ mag, which is likely
too high for known stars.  If Star 1 is not the progenitor, which we consider
more likely, then $M^0_V \gtrsim -8.7$, $(U-B)^0{\lesssim}-0.3$,
$(B-V)^0{\lesssim}0.2$, and $(V-I)^0{\lesssim}0.9$ mag for the progenitor.

\subsection{SN 1999ev in NGC 4274}

SN 1999ev was discovered by T.~Boles of the U.K. Nova/Supernova Patrol (Hurst
1999) and was classified as a SN II by Garnavich et al.~(1999b).  We measure
from a KAIT image the SN position as $\alpha$ = 12$^h$19$^m$48${\fs}$20,
$\delta$ = +29{\arcdeg}37{\arcmin}21${\farcs}$7, with uncertainty
${\pm}0{\farcs}4$. This agrees, to within the errors, with the position
measured by Armstrong (1999), but disagrees with the positions measured by
Boles and by Garnavich et al.  The host galaxy was imaged by GO-5741 in F555W
(280 s) on 1995 February 5.  We applied the 2MASS grid directly to the WFPC2
mosaic for the four faint stars in common, with uncertainty ${\pm}0{\farcs}7$,
leading to a total uncertainty ${\pm}0{\farcs}8$.  Figure 10 shows the SN site
on the WF2 chip.  The two brightest of the faint objects within the error
circle, labelled A and B, have $m_{\rm F555W}=24.93{\pm}0.23$ and
$25.28{\pm}0.32$ mag, respectively.  Assuming a distance of about 17 Mpc and
Galactic $A_V=0.07$ mag, these correspond to $M^0_V \approx -6.3$ and $-5.9$
mag, respectively.  Either of these is consistent with the absolute magnitudes
of red supergiants and could be the progenitor.  If neither of these two stars
is the progenitor, then it had $M^0_V \gtrsim -5.5$ mag.

\subsection{SN 2000C in NGC 2415}

SN 2000C was independently discovered by S. Foulkes and M. Migliardi (Foulkes
et al. 2000). It was classified as a SN~Ic by Cappellaro et al.~(2000) and Jha
et al.~(2000).  We measure the SN position from a galaxy background-subtracted
KAIT image as $\alpha$ = 07$^h$36$^m$57${\fs}$11s, $\delta$ =
+35{\arcdeg}14{\arcmin}39${\farcs}$0, with uncertainty ${\pm}0{\farcs}2$.  This
position agrees, to within the errors, more with that measured by Migliardi
than with that measured originally by Li.

We (GO-8602) obtained a F555W 700-s Snapshot image (in a cosmic-ray split pair)
on 2001 March 11.  The host galaxy was also imaged by GO-6862 on 1997 May 19 in
F547M (700 s), F656N (600 s), and F814W (560 s), and by GO-9124 on 2002 May 3
in F300W (600 s).  We applied the 2MASS astrometric grid directly to the
unsaturated and partially saturated stars on the F555W mosaic, with a total
uncertainty of ${\pm}0{\farcs}5$ in the SN position on the mosaic.  Figure 11
shows this position on the PC chip.  Within the error circle is a point source
with relatively high signal-to-noise ratio (S/N), which is almost certainly the
SN at late times, with $m_{\rm F555W}=22.74{\pm}0.07$ mag.  We can be confident
of the SN identification, since this object is seen on both of the cosmic-ray
splits {\it and\/} is not seen in the pre-SN F547M image of similar depth
(Figure 12) or in the F814W PC images.  The SN is not detected on the F300W WF3
chip on 2002 May 3 to $m_{\rm F300W}\gtrsim 22.9$ mag (however, the star
cluster to the west of the SN is quite bright in this band).  In neither the
F547M nor F814W image is a progenitor candidate detected, to $V \gtrsim 25.1$
and $V-I \lesssim 1.3$ mag.  For a distance of about 60 Mpc and Galactic
$A_V=0.14$ mag, this corresponds to $M^0_V \gtrsim -8.9$ and $(V-I)^0 \lesssim
1.2$ mag for a progenitor.

SN 1998Y, which was discovered by Li et al.~(1998) using KAIT and classified as
a SN~II by Filippenko, Leonard, \& Riess (1998), also occurred in this host
galaxy. Because of poor seeing, the faintness of the SN, and high galactic
background in the KAIT images, it is difficult to obtain a good centroid for
the SN in order to accurately determine its position.  From a relatively crude
position and its proximity to SN 2000C (within 2\arcsec), we know that the SN
site is also on the Snapshot image (PC chip).  We have very carefully
subtracted the light of the pre-SN F547M image from that of the Snapshot image
and inspected a circular region with radius 60 pixels centered on the SN 2000C
position (60 pixels on the PC
chip is $\sim$3\arcsec), but we could not locate SN 1998Y.  It must have faded
below detectability by 2001 March.  (It should be noted that the subtraction
very nicely reveals SN 2000C.)  The limit on the brightness of SN 1998Y in 2001
March is $V \gtrsim 25.9$ mag.  We did not try to locate the SN 1998Y
progenitor in any of the pre-SN images, owing to the crude position.

\subsection{SN 2000ds in NGC 2768}

SN 2000ds was discovered by Puckett \& Dowdle (2000) and classified as a
relatively old SN Ib by Filippenko \& Chornock (2000).  From a KAIT image we
measure the SN position as $\alpha$ = 09$^h$11$^m$36${\fs}$28, $\delta$ =
+60{\arcdeg}01{\arcmin}43${\farcs}$3, with uncertainty ${\pm}0{\farcs}3$.  The
host galaxy was imaged by GO-6587 on 1999 May 20 in F555W (1000 s and also 400
s) and in F814W (2000~s).  Only two of the stars seen on the 1000-s F555W
mosaic have 2MASS counterparts; the rest are too faint.  For this reason we
established a secondary astrometric grid using a deep Palomar $V$-band image of
the host, with uncertainty ${\pm}0{\farcs}4$.  The resulting grid applied to
the WFPC2 mosaic has uncertainty ${\pm}0{\farcs}5$, for a total uncertainty of
${\pm}0{\farcs}7$ in the SN position on the mosaic.  Figure 13 shows the SN
site on the WF2 chip.  Although hints of faint, reddish clusters of stars may
be evident near the SN position at the 3--4$\sigma$ level ($I {\approx}
25.7$--24.7 mag), generally no star is detected to $V \gtrsim 26.6$ and $V-I
\lesssim 1.6$ mag.  For a distance of about 25 Mpc and Galactic $A_V=0.15$ mag,
this corresponds to $M^0_V \gtrsim -5.5$ and $(V-I)^0 \lesssim 1.5$ mag for the
progenitor.

\subsection{SN 2000ew in NGC 3810}

SN 2000ew was discovered by Puckett \& Langoussis (2000) and classified as a SN
Ic by Filippenko, Chornock, \& Modjaz (2000).  From a KAIT image we measure the
SN position as $\alpha$ = 11$^h$40$^m$58${\fs}$60, $\delta$ =
+11{\arcdeg}27{\arcmin}55${\farcs}$8, with uncertainty ${\pm}0{\farcs}3$.
[This position differs by $\sim$1\arcsec\ or more from those reported by
Puckett \& Langoussis and by Garradd (2000).]  The host galaxy was imaged by
GO-9042 in F450W and F814 W (460 s each) on 2001 November 7--8 and by GO-5446
in F606W (160 s) on 1994 November 4.  The host is quite large on the mosaics,
and the Galactic stars on the F814W mosaic are too faint for 2MASS, so we
established an astrometric grid using a deep Palomar $V$-band image.  Applying
this grid to the WFPC2 mosaic, the uncertainty is ${\pm}0{\farcs}6$, leading to
a total uncertainty of ${\pm}0{\farcs}7$ in the SN position on the mosaics.

In Figure 14 we show the SN position on the F814W WF3 chip.  Toward the
southeast edge of the error circle is a point source, which we identify as the
fading SN. Our confidence stems primarily from the fact that the point source
is seen in both the F450W and F814W (2001 November 7--8) images, {\it and\/} it
is not seen in the F606W (1994 November 4) WF4 image (Figure 15).  Note how
close the position is to the chip edge ($<$50 pixels).  Since HSTPhot masks the
first 50 pixels from the chip edge, in this exceptional case we had to use PSF
fitting in DAOPHOT/ALLSTAR (Stetson 1987, 1992) within IRAF, with a Tiny Tim
PSF (Krist 1995) for both the F450W and F814W bands, and subsequently tie the
results to the HSTPhot output for point sources across the rest of the unmasked
chip.  Finally, we transformed the resulting magnitudes to standard $B$ and $I$
via synthetic photometry generated with the STSDAS package SYNPHOT and the
Bruzual Spectral Atlas (see Filippenko et al.~1995; Van Dyk et al.~2002a).  We
estimate that the SN had $B=22.78{\pm}0.05$ and $I=20.97{\pm}0.04$ mag on 2001
November 8.

A progenitor is not detected on the pre-SN image, to $m_{\rm
F606W}{\gtrsim}24.7$ mag.  The Galactic $A_V=0.15$ mag, but we can use the
F450W and F814W image color information to investigate the extinction local to
the SN.  Figure 16 illustrates the color-magnitude diagram for the SN
environment, showing the SN and the three stars immediately next to the SN in
Figure 14.  The diagram implies that all three stars are quite luminous,
intermediate in color, and young.  The three stars could all be in the
helium-burning phase, following the blue loop expected for massive star
evolution, or they could all be very blue supergiant stars experiencing $A_V
\approx 1$--1.5 mag.  The fact that the environment appears to contain blue or
yellow, young ($\lesssim 6$ Myr), and therefore massive, stars suggests that
the SN progenitor could have been quite massive as well.  For a distance of
about 16 Mpc and assuming simply the Galactic value of $A_V$, a progenitor had
$M^0_V \gtrsim -6.5$ mag.  For the possibly larger extinction range, this
corresponds to $M^0_V \gtrsim -7.3$ to $-7.8$ mag.  We clearly require more
color information to better understand the SN's environment and its possible
progenitor.

\subsection{SN 2001B in IC 391}

SN 2001B was discovered by the BAO SN search (Xu \& Qiu 2001) and classified as
a probable SN Ib by Chornock \& Filippenko [2001; note that Matheson et
al.~(2001b) had earlier classified it as SN~Ia].  From a Mt.~Hopkins $V$-band
image (kindly provided by T.~Matheson) we measure the SN position as $\alpha$ =
04$^h$57$^m$19${\fs}$31, $\delta$ = +78{\arcdeg}11{\arcmin}16${\farcs}$6, with
uncertainty ${\pm}0{\farcs}2$.  The host galaxy was imaged by GO-5104 in a
single 70-s F555W exposure on 1994 February 21.  We use four stars from the
Mt.~Hopkins image to establish the astrometric grid on the WFPC2 mosaic, with
total uncertainty ${\pm}0{\farcs}3$.  Figure 17 shows the SN site on the WF3
chip.  Within the error circle is a point source with $m_{\rm
F555W}=23.38{\pm}0.18$ mag.  For a distance of about 28 Mpc and Galactic
$A_V=0.42$ mag, we find $M^0_V {\approx} -9.3$ mag for this star, which we
tentatively identify as the SN progenitor.  Unfortunately, we do not have any
color information for the candidate.  However, if this star is not the
progenitor, then for a detection limit $m_{\rm F555W}\gtrsim 24.3$ mag, the
progenitor had $M^0_V \gtrsim -8.4$ mag.

\subsection{SN 2001ai in NGC 5278}

SN 2001ai was discovered by LOTOSS (Modjaz, Li, \& Schwartz 2001a) and
classified as a SN~Ic by Matheson et al.~(2001c).  From a KAIT image we measure
the SN position as $\alpha$ = 13$^h$41$^m$39${\fs}$37, $\delta$ =
+55{\arcdeg}40{\arcmin}05${\farcs}$8, with uncertainty ${\pm}0{\farcs}3$.  The
host galaxy was imaged in F255W, F300W, and F814W (total exposure times 1400 s,
1500 s, and 260 s, respectively) on 2000 December 18 by GO-8645.  Since the
stars seen in the F814W mosaic are too faint for 2MASS counterparts, we
establish the astrometric grid from a Palomar $V$-band image, with uncertainty
${\pm}0{\farcs}3$.  We show in Figures 18 and 19 the SN site on the WF3 chip,
with total uncertainty ${\pm}0{\farcs}4$, in F814W and F300W, respectively.

Two objects which HSTphot considers extended, A and B, are detected in both
F300W and F814W at the periphery of the error circle, with $m_{\rm F300W} =
21.10{\pm}0.01$, $m_{\rm F814W} = 21.35{\pm}0.01$ mag, and $m_{\rm F300W} =
20.96{\pm}0.01$, $m_{\rm F814W} = 22.14{\pm}0.01$ mag, respectively.
Additionally, Object C, considered stellar by HSTphot, is detected at F300W
only at the southern edge of the circle, with $m_{\rm F300W} = 22.57{\pm}0.32$
mag.  (The F255W image is not of sufficiently high S/N to provide additional
information on the environment; the detection limit is $m_{\rm F255W} \gtrsim
20.5$ mag.)  No other objects are detected within the error circle to $m_{\rm
F300W} \gtrsim 23.2$ and $m_{\rm F814W} \gtrsim 23.9$ mag.

The SN host is quite distant ($\sim$120 Mpc), and for Galactic $A_V=0.03$ mag,
Objects A and B have $M^0_I\approx -14.0$, $(U-I)^0 \approx -0.3$ mag, and
$M^0_I\approx -13.3$, $(U-I)^0 \approx -1.2$ mag, respectively. Object C has
$M_U\approx -12.8$, $(U-I)^0 \lesssim -1.3$ mag (based on the $m_{\rm F814W}$
detection limit), which is likely too bright for a single star.  At 120 Mpc,
in this case what HSTphot considers stellar is probably still an extended
object (e.g., a compact cluster).  The progenitor could have been associated
with one of these three objects, which are probably star clusters, and was not
resolved.  Alternatively, the progenitor was not associated with these objects
and also not detected, to $M^0_I \gtrsim -11.5$ mag, the upper end of which
greatly exceeds even the most luminous supergiants.  The relatively
unrestrictive color limit for an undetected progenitor is $(U-I)^0 \lesssim
-0.3$ mag.

\subsection{SN 2001ci in NGC 3079}

SN 2001ci was discovered on 2001 April 25 by LOTOSS using KAIT (Swift, Li, \&
Filippenko 2001).  Swift et al.~remarked on the low apparent luminosity of the
SN.  Filippenko \& Chornock (2001) identified it as a SN Ic, but possibly
extinguished by $A_V \approx 5$--6 mag.  We measure from a KAIT image a
position of $\alpha$ = 10$^h$01$^m$57${\fs}$21, $\delta$ =
+55{\arcdeg}41{\arcmin}14${\farcs}$0, with uncertainty ${\pm}0{\farcs}3$.  The
host galaxy was imaged on 2001 January 21 in F606W (560 s) by GO-8597, and on
1999 March 4 by GO-7278 in F547M (320 s) and F814W (140 s), all pre-SN
observations.  The stars on the F606W mosaic were too faint for 2MASS, so we
used a deep Palomar $V$-band image to establish the astrometric grid, with
uncertainty ${\pm}0{\farcs}2$, and total uncertainty ${\pm}0{\farcs}3$.  The
host was also imaged on 2001 December 9, well after discovery, in F300W
(${\sim}U$; 800 s) by GO-9124, but the SN is not detected to $m_{\rm F300W}
\gtrsim 23.5$ mag.  We show in Figure 20 the SN site on the F814W WF3 chip.  A
hint of an object can be seen within the error circle, but no star is detected
by HSTphot to $V \gtrsim 24.7$, $V-I \lesssim 1.8$ mag, based on the F547M and
F814W images. The F606W limit is significantly deeper, at $V \gtrsim 26.4$ mag.
Assuming the above range in extinction and a distance of about 21 Mpc, the
F606W limit corresponds to $M^0_V \gtrsim -10.2$ to $-11.2$ mag.  The color
limit is $(V-I)^0 \lesssim -0.2$ to $-0.6$ mag.

\subsection{SN 2001du in NGC 1365}

SN 2001du was visually discovered by Evans (2001), about 90\arcsec\ W and
10\arcsec\ S of the nucleus of the nearby barred spiral galaxy NGC 1365.  The
SN was classified as Type II-P by Wang et al.~(2001).  We have independently
measured the position from three different SN images available on the Internet:
an image by G.~Bock, a deeper image by T.~Dobosz, and an image obtained with
the YALO 1-m at CTIO.  We derive three slightly different positions,
respectively: $\alpha$ = 3$^h$33$^m$29${\fs}$15, $\delta$ =
$-$36{\arcdeg}08{\arcmin}32${\farcs}$0; $\alpha$ = 3$^h$33$^m$29${\fs}$14,
$\delta$ = $-$36{\arcdeg}08{\arcmin}32${\farcs}$0; and, $\alpha$ =
3$^h$33$^m$29${\fs}$15, $\delta$ = $-$36{\arcdeg}08{\arcmin}31${\farcs}$5.
(Uncertainties in each are ${\sim}0{\farcs}7$, ${\sim}0{\farcs}5$, and
${\sim}0{\farcs}3$, respectively.)  Together, these measurements differ from
each other by ${\sim}0{\farcs}7$.  The measurements all differ by
${\gtrsim}0{\farcs}5$ from that measured by Jacques (2001).  We adopt the
position measured from the YALO image, with the smallest error.  This adopted
position is ${\sim}0{\farcs}7$ from the position we initially quoted in Van Dyk
et al.~(2001); the difference is likely due to the improved astrometric
solution, but is consistent with the overall uncertainties discussed here.
Given the uncertainty in the measured positions, and the relative disagreement
with previously measured positions, we adopt a total positional uncertainty of
${\pm}0{\farcs}8$.

The SN site is in 100-s F336W, F555W, and F814W exposures obtained by GTO-5222
on 1995 January 15. [Unfortunately, the Cepheid Key Project (Silbermann et
al. 1999) very deep images are of the opposite arm of NGC 1365.] We could apply
the 2MASS astrometric grid using only three relatively bright stars on the
F555W mosaic.  We estimate the uncertainty in this application, based on one
other object on the mosaic, and find ${\pm}0{\farcs}4$, which leads to a grand
total uncertainty of ${\pm}0{\farcs}9$.  Figure 21 shows the SN site on the
F555W WF3 chip.  Three objects, A--C, are detected within the error circle (a
source that looks somewhat extended is toward the center of the circle, but it
is undetected by HSTphot).  Object A, to the south, is blue, with $m_{\rm
F555W} = 24.30 {\pm} 0.21$ mag and no detection at F814W; Object B, to the
west, is also blue, with $m_{\rm F555W} = 25.02 {\pm} 0.32$ mag and no F814W
counterpart; and Object C, to the east, is relatively red, with $V = 24.44
{\pm} 0.23$ and $V-I = 1.03{\pm}0.30$ mag.  (The F336W exposures are of
insufficient S/N to show any object at the SN site.)  Since SN 2001du is of
Type II-P, we assume that of the three detected candidates, this red eastern
star is the most plausible progenitor (but still unlikely; see below).

The Galactic extinction toward the host is $A_V=0.07$ mag, but we possibly can
use the color information from the F555W and F814W images to estimate the
extinction local to the SN, as well as study the properties of the SN's stellar
environment.  Figure 22 shows the ($V-I$, $V$) color-magnitude diagram for the
environment.  The reddish progenitor candidate, Star C, is indicated.  It is
interesting that no red (M-type) supergiant stars, the presumed SN~II
progenitors, with $V-I {\gtrsim} 1.8$ and $V {\gtrsim} 25$ mag, are detected in
the environment, likely due to the low S/N of these images.  It is also notable
that many of the detected stars, including Star C, have $V \approx 24.7$ and
$V-I \approx 1.1$ mag.  These presumably K-type supergiants either all have
ages $\sim$12--16 Myr and are in the blue loop core He-burning phase, or they
are intrinsically far bluer and younger supergiants experiencing similar,
larger amounts of extinction, $A_V \approx 2.5$ mag.

Given the distance modulus $\mu = 31.3$ mag to NGC 1365 determined from {\sl
HST\/} observations of Cepheids (Silbermann et al.~1999) and only the Galactic
extinction, if Star C is the progenitor it has $M^0_V \approx -6.9$ and
$(V-I)^0 \approx 1.0$ mag.  With possibly higher local extinction, $A_V \approx
2.5$ mag, this becomes $M^0_V \approx -9.4$ and $(V-I)^0 \approx 0.0$ mag.  We
consider it more likely that the relevant archival data were just not sensitive
enough to detect the true progenitor of SN 2001du.  Assuming only Galactic
extinction, the progenitor had $M^0_V \gtrsim -6.3$ and $(V-I)^0 \lesssim 1.5$
mag.  Program GO-9041 has imaged SN 2001du in several bands with WFPC2, and
these data will be public in late November 2002.  At that time, or possibly
earlier, we will know the exact location of the SN and potentially learn more
about the nature of the progenitor.

\subsection{SN 2001is in NGC 1961}

SN 2001is was independently discovered by both the BAO and LOTOSS searches (Qiu
et al.~2001).  Benetti et al.~(2001) identified it as a SN Ib, with possible
residual H contamination.  From a KAIT image we measure the SN position as
$\alpha$ = 05$^h$42$^m$09${\fs}$12, $\delta$ =
+69{\arcdeg}21{\arcmin}54${\farcs}$5, with uncertainty ${\pm}0{\farcs}4$.  The
host galaxy was imaged (GO-5419) on 1994 August 28 in a cosmic-ray split F218W
exposure (1800~s total) and in a single 300-s F547M exposure.  The host was
also imaged deeply by GO-9106 in the FR680N and F547M bands (4000~s each) on
2001 July 14.  (The low S/N of the FR680N images near the SN site does not make
them particularly useful.)  The KAIT image was sufficiently deep that we could
identify some of the stars in each of the F547M mosaics, with astrometric
uncertainty ${\pm}0{\farcs}2$, for a total uncertainty of ${\pm}0{\farcs}4$ in
the SN's position on the mosaics.  In Figure 23 we show the SN site on the deep
F547M WF3 chip.  Two stars, A and B, are detected within at $m_{\rm
F547M}=26.11{\pm}0.27$ and $25.77{\pm}0.20$ mag.  Assuming a distance of about
64 Mpc and Galactic $A_V=0.41$ mag, we find $M^0_V\approx -8.3$ and $-8.7$ mag
for the two stars, respectively, either of which could be the progenitor. If
neither of these stars is the progenitor, then for a detection limit of $m_{\rm
F547M}\gtrsim 26.4$ mag, the progenitor had $M^0_V\gtrsim -8.0$ mag.

\section{Discussion}

In Table 2 we summarize the results of our search for the progenitors of
core-collapse SNe.  For SNe with candidate progenitor identifications, we have
also included, in parentheses, the limits on the absolute magnitude and color
of a possible progenitor, if the candidate we have identified is not the actual
progenitor.  It should be noted that for all the magnitude and color estimates,
we have assumed possibly inaccurate distance estimates and, in most cases, only
the Galactic component to the extinction toward the SN.  Consequently, we may
have either underestimated or overestimated the absolute magnitude, or limits
on the absolute magnitude, of candidate progenitors.  Extinction local to the
SNe within the host galaxies themselves will only lead to higher absolute
brightnesses, and to bluer intrinsic colors, or color limits, for all these
objects.  At the least, we provide the measured magnitudes and magnitude limits
for the SN progenitors, based on the HSTphot output, so that, with additional
information, the reader can make his or her own estimations.  Below, we briefly
interpret our results, but we eschew transforming the observed magnitudes and
colors into intrinsic properties, such as bolometric luminosity and surface
temperature, for lack of adequate information about the stars.

The candidate SNe II progenitors of SNe 1999br, 1999ev, and 2001du all have
absolute magnitudes ($M^0_V \approx -5.9$ to $-6.9$) that are consistent with
the known red supergiants in the Galaxy.  All of the magnitude limits for the
other SNe~II are also consistent with supergiant stars.  The only candidate
with intrinsic color information, the progenitor candidate for SN 2001du, has a
$(V-I)^0$ value more consistent with an early K spectral type (e.g., the
Vilnius spectral colors in Bessell 1990), rather than the M-type supergiant
that we would expect, based on theoretical models.  In fact, even if this
candidate is not the progenitor, the $(V-I)^0$ color limit for the SN 2001du
progenitor, as well as the color limits for the SN 1999bx progenitor [if we
discount the $(U-V)^0$ limit], imply spectral types that can only be as late as
early M-type.  The slightly bluer color might imply a somewhat more compact
morphology for the progenitor envelope, or possible contamination of its light
by a close, fainter and bluer companion star in a possible binary system.

Models for SNe~Ib include the explosion of isolated Wolf-Rayet stars or of
helium stars in binary systems, possibly with a wide orbit including a more
massive main-sequence secondary (e.g., van den Heuvel 1994).  Both mechanisms
could lead to SNe~Ib.  However, Branch et al.~(2002) find, from their analysis
of SN~Ib spectral data, that the masses and kinetic energies among SNe~Ib are
similar, implying that progenitor masses must be similar as well.  Wolf-Rayet
stars that have been stripped of much of their helium could lead to SNe~Ic;
however, model light curves decline too slowly, compared to observations
(Woosley, Langer, \& Weaver 1993).  Low-mass helium stars in binaries (e.g.,
Nomoto et al.~1990) end up with too much helium for SNe~Ic.  Nomoto et
al.~(1994) explored a $\sim$2--3 $M_\odot$ C+O star (which originally evolved
from a 13--18 $M_\odot$ main-sequence star) in a close binary as the progenitor
of the SN~Ic 1994I in M51.  Their model, possibly including a common-envelope
phase, involves two episodes of mass transfer with a secondary, which
eventually evolves to be a low-mass main sequence star, a neutron star, or a
white dwarf.  All of these scenarios involve compact stars, which are likely
not particularly luminous; Barth et al.~(1996) place an upper limit of $M_V
{\gtrsim} -7.3$ mag on the SN 1994I progenitor.  As Podsiadlowski et al.~(1992)
point out, mass transfer, in fact, could make the secondary more massive and
more luminous when the primary explodes, meaning that a detected ``progenitor''
could actually be the companion star.

The candidate progenitor for the SN Ic 1999bu is quite luminous, with
$M^0_V \approx -7.5$ mag, which is consistent with the upper limit for the 
SN 1994I progenitor mentioned above.
The candidate progenitors for the SNe Ib 2001B and 2001is are also very luminous,
with $M^0_V \approx -8$ to $-9$ mag.  The known Wolf-Rayet stars in the Galaxy
have absolute magnitudes spread over a large range, $M^0_V
\approx -2$ to $-8$ mag (van der Hucht 2001).  The SN Ib progenitor candidate
luminosities are near or slightly above the upper end of this luminosity range.
Therefore, it is possible that these two SNe Ib arose from Wolf-Rayet stars.
(However, with the uncertainties in the distances to and reddening within the
host galaxies, these absolute magnitudes may actually fall outside the
Wolf-Rayet luminosity range, i.e., they would be too bright.)  The luminosities
for these SN Ib progenitor candidates are also consistent with those of other, less
evolved, presumably blue or yellow supergiants (Humphreys \& Davidson 1979).
Alternatively, it is possible that these candidates are not the progenitors,
but instead multiple star systems or compact star clusters.

Generally, the absolute magnitude limits for all other SNe~Ib/c (the brightness
limits for SNe 2001ai and 2001ci are not very restrictive) are consistent both
with the range of Wolf-Rayet magnitudes and also with expectations for
interacting binary models.  The intrinsic color limits, although also usually
not very restrictive, are consistent with blue or yellow stars.  The most
restrictive color limits are for the progenitors of SNe 1999ec and 2001ci (if,
in this case, the extinction estimate $A_V \approx 5$--6 mag is, in fact,
correct): taken together, these limits imply that the progenitors had spectral
types of A-type or earlier (Bessell 1990) and are also consistent with the
range of $B-V$ (about $-0.5$ to $-0.1$ mag) for Galactic Wolf-Rayet stars (van
der Hucht 2001).  The host galaxy of SN 2001ci, NGC 3079, is seen nearly
edge-on, so it is possible that the SN progenitor could have been a Wolf-Rayet
star exploding while obscured by or embedded in dust.

To some extent, our search was not fully satisfying.  Although it is true that,
compared with the case of Van Dyk et al.~(1999b), the {\sl HST\/} archive is
currently much richer in pre-SN host galaxy images, which may potentially
contain SN progenitors, and more importantly, post-SN images in which the
fading SNe may be recovered, the quality of these data is such that the S/N or
number of filters used are generally still not ideal.  The data in our sample
are not yet sensitive enough for us to place more stringent constraints on the
competing models for SNe~Ib/c.  The {\sl HST\/} archive will steadily grow,
however, as observations continue after the recent refurbishment mission.
These new observations will also include images of galaxies with the superior
Advanced Camera for Surveys (ACS).  Additionally, new nearby SNe will be
discovered by LOSS/LOTOSS and other SN searches, greatly expanding the
potential sample.  We intend to further exploit the {\sl HST\/} archive to
continue to search for core-collapse SN progenitors; the future looks bright
for this subject.

\section{Conclusions}

We have searched for the progenitor stars of 16 core-collapse SNe using
archival {\sl HST\/} WFPC2 images.  The sample includes 6 SNe~II and 10
SNe~Ib/c.  We may have identified the progenitors of the SNe II 1999br, 1999ev,
and 2001du as supergiant stars with $M^0_V \approx -6$ mag in all three cases.
We may also have identified the progenitors of the SNe Ib 2001B and 2001is as
very luminous supergiants with $M^0_V \approx -8$ to $-9$ mag, and possibly the
progenitor of the SN Ic 1999bu as a supergiant with $M^0_V\approx -7.5$ mag.
If these identifications can be verified, this more than doubles the number of
known SN progenitors from five to eleven.  For all other SNe in our sample we
could only place limits on the progenitor absolute magnitude and color (when
multi-band images were available).  We have also recovered SNe 1999dn, 2000C,
and 2000ew at late times.  Unfortunately, the pre-SN images for these recovered
SNe did not show a progenitor candidate at the SN position.

The possible detections and constraints on the SN II progenitors are broadly
consistent with red supergiants as progenitor stars, although the progenitor
candidates are not as red as would be expected, with their colors implying
spectral types typically earlier than M.  The SN~Ib progenitor candidates may
well be Wolf-Rayet stars, although possibly at the upper luminosity end for
known stars of this kind.  The lone SN Ic progenitor candidate is consistent
with a luminous supergiant star.  In general, we cannot place rigorous constraints 
on either the Wolf-Rayet star or massive interacting binary models for SN~Ib/c
progenitors, based on these data.  For both the SNe~II and Ib/c uncertainties
in the host galaxy distances and extinction toward the SNe also limit what
conclusions we can draw about the progenitor stars.  However, from purely
environmental considerations, our results are consistent with those of Van Dyk
et al.~(1999b), who found that SNe~Ib/c seem to be more closely associated with
massive stellar regions than is true for SNe~II; five of the SNe~Ib/c in our
sample (SNe 1999ec, 2000C, 2000ew, 2001ai, and 2001is) occurred very near
bright, possibly blue and young star clusters (the cluster near SN 2000C, based
on the F300W image, is quite blue, and the cluster near SN 2000ew, based on the
color-magnitude diagram, is quite young).  Again, the statistics are still
small, but this continues to suggest that at least some SN~Ib/c progenitors may
be more massive, in general, than SN~II progenitors, consistent with the
Wolf-Rayet model.

A current program (GO-9353) is imaging six of the SNe in our sample (1999an,
1999br, 1999ev, 2000ds, 2000ew, and 2001B) in multiple bands with ACS, to
attempt to recover them at late times, with the same aim of matching the pre-SN
and post-SN images to identify the SN progenitor. Similarly, program GO-9041
has imaged SN 2001du with WFPC2, but the data are not yet public.  For SNe
1999an, 2000ds, and 2000ew, we have already likely recovered SN 2000ew, and
we have already shown here that the pre-SN images
are simply not of sufficient S/N to detect each SN progenitor.  For SNe 1999br,
1999ev, 2001B, and 2001du, the new observations from these programs will be
quite revealing: if the SNe are actually recovered, they may or may not match
up positionally with the progenitor candidates that we have identified.

In this paper we have made some progress toward a statistically significant
sample of core-collapse SN progenitors directly identified on {\sl HST\/} image
data.  However, the limitations of the data continue to be the restricted field
of view, low S/N, and poor color coverage.  With the full operation of the
newly commissioned ACS onboard {\sl HST}, and with additional galaxies observed
using WFPC2 as well, the amount of available archive data will continue to
grow, providing for larger SN samples in the future.

\acknowledgements
This publication makes use of data products from the Two Micron All Sky Survey,
which is a joint project of the University of Massachusetts and the
IPAC/California Institute of Technology, funded by NASA and NSF.  This research
has also made use of the NASA/IPAC Extragalactic Database (NED) which is
operated by the Jet Propulsion Laboratory, California Institute of Technology,
under contract with NASA, and the LEDA database (http://leda.univ-lyon.fr).
The work of A.V.F.'s group at UC Berkeley is supported by NSF grant
AST-9987438, by the Sylvia and Jim Katzman Foundation, and by NASA grants
AR-8754, AR-9529, and GO-8602 from the Space Telescope Science Institute, which
is operated by AURA, Inc., under NASA contract NAS5-26555.  KAIT was made
possible by generous donations from Sun Microsystems, Inc., the Hewlett-Packard
Company, AutoScope Corporation, Lick Observatory, the National Science
Foundation, the University of California, and the Katzman Foundation.  We thank
A.~J.~Barth and D.~C.~Leonard for useful discussions.

\clearpage

\begin{deluxetable}{lllccl}
\def\phmm{\phm{$-$}}
\tablenum{1}
\tablecolumns{6}
\tablecaption{Summary of Available Data}
\tablehead{\colhead{SN} & \colhead{Host} & \colhead{Date} & \colhead{Filters}
& \colhead{Exp.~Time} & \colhead{{\sl HST}} \nl
\colhead{} & \colhead{Galaxy} & \colhead{(UT)} & \colhead{}
& \colhead{(s)} & \colhead{Program}}
\startdata
1998Y  & NGC 2415 & 1997 May 19 & F547M & 700 & GO-6862 \nl
          &          &             & F656N & 600 &         \nl
          &          &             & F814W & 560 &         \nl
          &          & 2001 Mar 11 & F555W & 700 & GO-8602\tablenotemark{a} \nl
          &          & 2002 May 03 & F300W & 600 & GO-9124 \nl
1999an & IC 755\phantom{}\phantom{} & 1995 Jan 05 & F606W & 160 & GO-5446 \nl
1999br & NGC 4900 & 1995 Jan 29 & F606W & 160 & GO-5446 \nl
          &          & 2002 Jun 20 & F450W & 460 & GO-9042 \nl
          &          &             & F814W & 460 &         \nl
1999bu & NGC 3786 & 1995 Mar 30 & F606W & 500 & GO-5479 \nl
1999bx & NGC 6745 & 1997 Mar 19,21 & F336W & 22000 & GO-6276 \nl
          &          &                & F555W &  4800 &         \nl
          &          &                & F675W &  5200 &         \nl
          &          &                & F814W &  5200 &         \nl
1999dn & NGC 7714 & 1996 May 15 & F606W & 500 & GO-5479 \nl
          &          & 1998 Aug 29 & F380W & 1800 & GO-6672 \nl
          &          & 2001 Jan 24 & F814W & 700 & GO-8602\tablenotemark{a} \nl
          &          & 2001 Jul 10 & F555W & 700 &         \nl
          &          &             & F814W & 700 &         \nl
          &          & 2001 Aug 03 & F300W & 300 & GO-9124 \nl
1999ec & NGC 2207 & 1996 May 25 & F336W & 2000 & GO-6483 \nl
          &          &             & F439W & 2000 &         \nl
          &          &             & F555W & 660 &         \nl
          &          &             & F439W & 720 &         \nl
1999ev & NGC 4274 & 1995 Feb 05 & F555W & 280 & GO-5741 \nl
2000C  & NGC 2415 & 1997 May 19 & F547M & 700 & GO-6862 \nl
          &          &             & F656N & 600 &         \nl
          &          &             & F814W & 560 &         \nl
          &          & 2001 Mar 11 & F555W & 700 & GO-8602\tablenotemark{a} \nl
          &          & 2002 May 03 & F300W & 600 & GO-9124 \nl
2000ds & NGC 2768 & 1999 May 20 & F555W & 400 & GO-6587 \nl
          &          &             &       & 1000 &        \nl
          &          &             & F814W & 2000 &        \nl
2000ew & NGC 3810 & 1994 Nov 04 & F606W & 160 & GO-5446 \nl
          &          & 2001 Nov 07,08 & F450W & 460 & GO-9042 \nl
          &          &                & F814W & 460 &         \nl
2001B  & IC 391\phantom{}\phantom{}   & 1994 Feb 21 & F555W & 70 & GO-5104 
\nl
2001ai & NGC 5278 & 2000 Dec 18 & F255W & 1400 & GO-8645 \nl
          &          &             & F300W & 1500 &         \nl
          &          &             & F814W & 260 &          \nl
2001ci & NGC 3079 & 1999 Mar 04 & F547M & 320 & GO-7278 \nl
          &          &             & F814W & 140 &         \nl
          &          & 2001 Jan 21 & F606W & 560 & GO-8597 \nl
\tablebreak
          &          & 2001 Dec 09 & F300W & 800 & GO-9124 \nl
2001du & NGC 1365 & 1995 Jan 15 & F336W & 100 & GTO-5222 \nl
          &          &             & F555W & 100 &          \nl
          &          &             & F814W & 100 &          \nl
2001is & NGC 1961 & 1994 Aug 28 & F218W & 1800 & GO-5419 \nl
          &          &             & F547M & 300 &          \nl
          &          & 2001 Jul 14 & F547M & 4000 & GO-9106 \nl
          &          &             & FR680N & 4000 &        \nl
\enddata
\tablenotetext{a}{This is part of our own Snapshot program (PI: Filippenko); see Li et 
al.~(2002).}
\end{deluxetable}

\clearpage

\begin{deluxetable}{lcc}
\def\phmm{\phm{$-$}}
\tablenum{2}
\tablecolumns{3}
\tablecaption{Summary of Progenitor Properties}
\tablehead{\colhead{SN} & \colhead{Absolute Magnitude\tablenotemark{a}}
& \colhead{Color\tablenotemark{a}} \nl
\colhead{} & \colhead{}
& \colhead{(mag)}}
\startdata
\cutinhead{SNe II}\nl
1998Y\tablenotemark{b} & \nodata & \nodata \nl
1999an & $M^0_V{\gtrsim}-6.9$          & \nodata             \nl
1999br & $M^0_V{\approx}-6.3$          & \nodata             \nl
          & ($M^0_V{\gtrsim}-5.9$)\tablenotemark{c}          & \nodata             \nl
1999bx & $M^0_V{\gtrsim}-7.8$            & $(U-V)^0{\lesssim}0.5$  \nl
          &                                 & $(V-R)^0{\lesssim}0.9$  \nl
          &                                 & $(R-I)^0{\lesssim}1.0$  \nl
          &                                 & $(V-I)^0{\lesssim}1.6$  \nl
1999ev & $M^0_V{\approx}-5.9$ to $-6.3$   & \nodata             \nl
          & ($M^0_V{\gtrsim}-5.5$)          & \nodata             \nl
2001du & $M^0_V{\approx}-6.9$             & $(V-I)^0{\approx}1.0$    \nl
          & ($M^0_V{\gtrsim}-6.3$)          & ($[V-I]^0{\lesssim}1.5$)  \nl
\cutinhead{SNe Ib/c}\nl
1999bu & $M^0_V{\approx}-7.5$             & \nodata             \nl
          & ($M^0_V{\gtrsim}-7.1$)          & \nodata             \nl
1999dn & $M^0_V{\gtrsim}-7.3$            & $(U-V)^0{\lesssim}2.5$  \nl
1999ec & $M^0_V{\gtrsim}-8.7$            & $(U-B)^0{\lesssim}-0.3$  \nl
          &                                 & $(B-V)^0{\lesssim}0.2$  \nl
          &                                 & $(V-I)^0{\lesssim}0.9$  \nl
2000C  & $M^0_V{\gtrsim}-8.9$            & $(V-I)^0{\lesssim}1.2$  \nl
2000ds & $M^0_V{\gtrsim}-5.5$            & $(V-I)^0{\lesssim}1.5$  \nl
2000ew & $M^0_V{\gtrsim}-6.5$            & \nodata             \nl
2001B  & $M^0_V{\approx}-9.3$             & \nodata             \nl
          & ($M^0_V{\gtrsim}-8.4$)          & \nodata             \nl
2001ai & $M^0_I{\gtrsim}-11.5$           & $(U-I)^0{\lesssim}-0.3$ \nl
2001ci & $M^0_V{\gtrsim}-10.2$ to $-11.2$ & $(V-I)^0{\lesssim}-0.2$ to $-0.6$ 
\nl
2001is & $M^0_V{\approx}-8.3$ to $-8.7$   & \nodata             \nl
          & ($M^0_V{\gtrsim}-8.0$)          & \nodata             \nl
\enddata

\tablenotetext{a}{The distance to the host galaxy is generally derived from the
heliocentric radial velocity corrected for Local Group infall into the Virgo
Cluster given in the LEDA database, with distance scale 65 km s$^{-1}$
Mpc$^{-1}$. The extinction and reddening to the SN are generally derived from
the Galactic component, adopted from NED, and assuming the Cardelli et
al.~(1989) reddening law.  See text for details.}
\tablenotetext{b}{We could not locate SN 1998Y in our own Snapshot (GO-8602)
images, nor could we determine a reliable position for the SN in the archival
images.}
\tablenotetext{c}{We have also included, in parentheses, the upper limits on
the absolute magnitude and color of a possible progenitor here and throughout
the table.}

\end{deluxetable}

\clearpage

\begin{figure}
\figurenum{1}
\caption{The site of SN 1999an in IC 755 in an archival 160-s F606W image from
1995 January 5.  The error circle has radius $0{\farcs}8$.  Six point-like
objects are detected within the circle, but with $M^0_V \approx -9.0$ to $-10.0$ mag,
they are somewhat too bright to be single stars and are more likely compact
star clusters.}
\end{figure}


\begin{figure}
\figurenum{2}
\caption{The site of SN 1999br in NGC 4900 in an archival 160-s F606W image
from 1995 January 29.  The error circle has radius $0{\farcs}5$.  A possible
progenitor star, with $M^0_V\approx -6.3$ mag, is identified along the
north-northeast edge of the circle and indicated with tick marks.}
\end{figure}


\begin{figure}
\figurenum{3}
\caption{The site of SN 1999br in NGC 4900 in an archival 460-s F814W image
from 2002 June 20. The error circle has radius $0{\farcs}5$.  The SN is not
recovered in this image to $I \gtrsim 24.5$ mag.}
\end{figure}


\begin{figure}
\figurenum{4}
\caption{The site of SN 1999bu in NGC 3786 in an archival 500-s F606W image
from 1995 March 30. The site was located by offsetting from a bright star to 
the northwest of the site, with an estimated error circle of radius $0{\farcs}5$.  
Three detected objects, A--C, are indicated with tickmarks.  Object C, with
$M^0_V \approx -7.5$ mag, could possibly be the progenitor of this SN Ic.}
\end{figure}


\begin{figure}
\figurenum{5}
\caption{The site of SN 1999bx in NGC 6745 in an archival 4800-s F555W image
from 1997 March 21.  The error circle has radius $0{\farcs}4$.  Four objects,
A--D, along the circle are indicated with tickmarks. They are probably too bright 
and too blue to be the progenitor star for this SN II.}
\end{figure}


\begin{figure}
\figurenum{6}
\caption{SN 1999dn in NGC 7714, as seen in our 700-s F814W Snapshot image
from 2001 January 24.  The SN is most likely the
star indicated with tickmarks within the $0{\farcs}6$
error circle.  It had $m_{\rm F814W}=24.18{\pm}0.18$ mag.
By July 10 the SN had faded below detectability in both the F555W and F814W
bands.  We have applied the routine {\it qzap\/} to remove residual
cosmic-ray hits.}
\end{figure}


\begin{figure}
\figurenum{7}
\caption{The site of SN 1999dn in NGC 7714 in a single archival 500-s F606W
exposure from 1996 May 15.  We have applied the routine {\it qzap\/} to make
the image more cosmetically appealing.  The error circle has radius
$0{\farcs}6$.  No star is detected at the exact position of the SN in Figure 6,
indicated with tickmarks.}
\end{figure}


\begin{figure}
\figurenum{8}
\caption{The site of SN 1999ec in NGC 2207 in an archival 660-s F555W
image from 1996 May 25.  The error circle has radius $0{\farcs}3$.
Two objects identified as stellar by HSTphot are indicated with tickmarks 
(see Figure 9).  Star 1, with $M^0_V \approx -11.2$ mag, is likely too 
bright to be a single star.  The progenitor, therefore, is likely not 
detected.}
\end{figure}


\begin{figure}
\figurenum{9}
\plotone{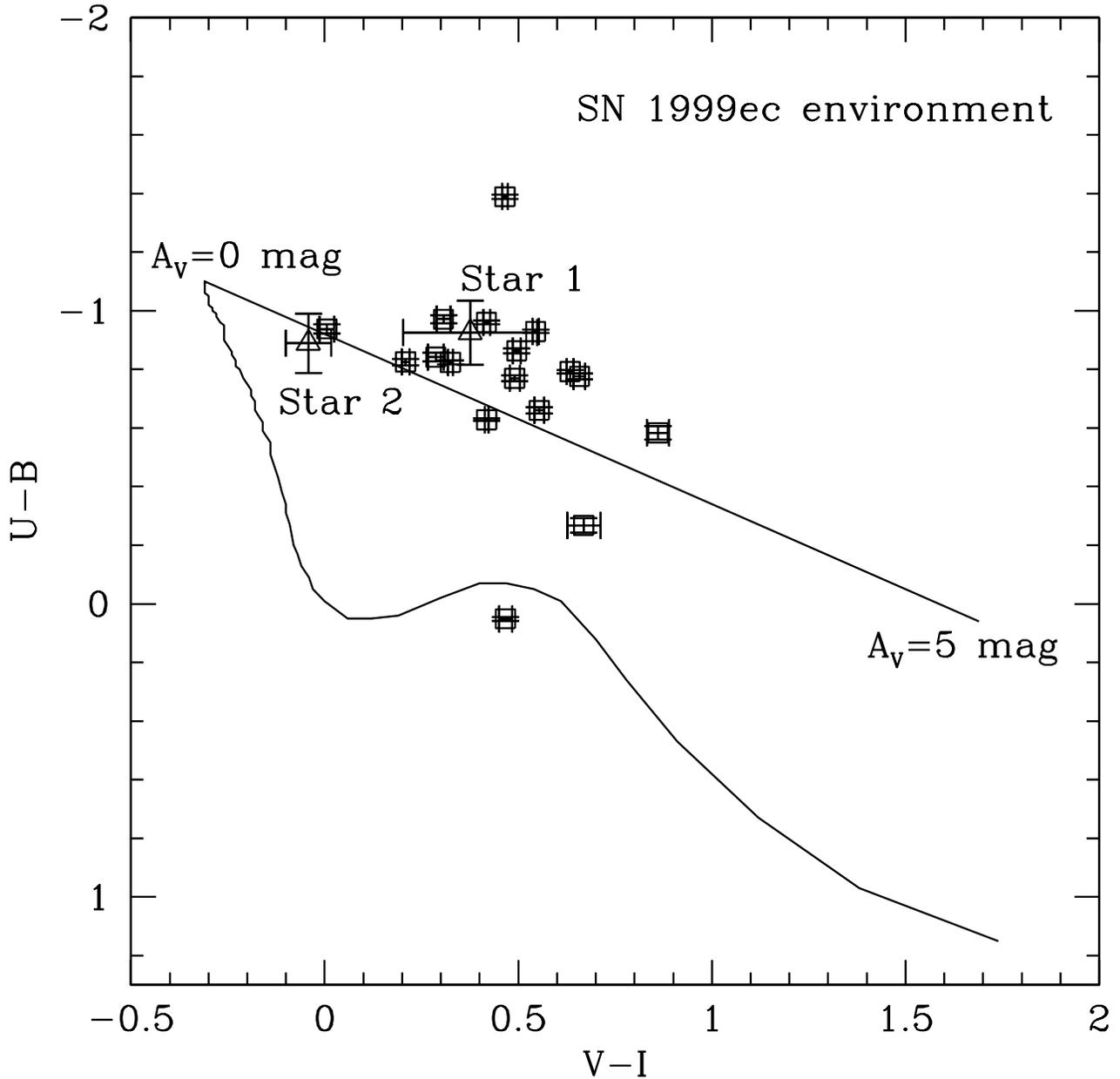}
\caption{The ($U-B$, $V-I$) color-color diagram for the SN 1999ec environment.
The two stellar objects indicated in Figure 8 are represented by {\it open 
triangles}; the extended objects are represented by {\it open squares}.  
The locus of the main sequence
is shown, as is the reddening vector, following the Cardelli et al.~(1989)
reddening law.}
\end{figure}


\begin{figure}
\figurenum{10}
\caption{The site of SN 1999ev in NGC 4274 in an archival 280-s F555W
image from 1995 February 5.  The error circle has radius $0{\farcs}8$.
Two faint objects within the error circle, A and B, with $M^0_V\approx -6.3$ 
and $-5.9$ mag, respectively, are indicated with
tickmarks.  Either of the two could be the progenitor of this SN II.}
\end{figure}


\begin{figure}
\figurenum{11}
\caption{The site of SN 2000C in NGC 2415 in a 700-s F555W Snapshot image
from 2001 March 11.  The error circle has radius $0{\farcs}5$.  The SN is
almost certainly the star with $m_{\rm F555W}=22.74$ mag, indicated with 
tickmarks within the circle.}
\end{figure}


\begin{figure}
\figurenum{12}
\caption{The site of SN 2000C in NGC 2415 in an archival 700-s F547M image
from 1997 May 19.
The error circle has radius $0{\farcs}5$.  No star is detected at the
exact position of the SN in Figure 10, indicated with tickmarks.}
\end{figure}


\begin{figure}
\figurenum{13}
\caption{The site of SN 2000ds in NGC 2768 in an archival 1000-s F555W image
from 1999 May 20.  The error circle has radius $0{\farcs}7$.  The progenitor
is not detected.}
\end{figure}


\begin{figure}
\figurenum{14}
\caption{The site of SN 2000ew in NGC 3810 in an archival 460-s F814W image
from 2001 November 8.  The error circle has radius $0{\farcs}7$.  The SN is
almost certainly the star, indicated with tickmarks, toward the southeast edge
of the circle.  It had $I=20.97$ and $B-I=1.81$ mag at that epoch.  Note how
close the site is to the edge of the WF3 chip.}
\end{figure}


\begin{figure}
\figurenum{15}
\caption{The site of SN 2000ew in NGC 3810 in an archival 160-s F606W image
from 1994 November 04.  The error circle has radius $0{\farcs}7$.  No star
is detected at the exact position of the SN in Figure 13, indicated with
tickmarks.}
\end{figure}


\begin{figure}
\figurenum{16}
\plotone{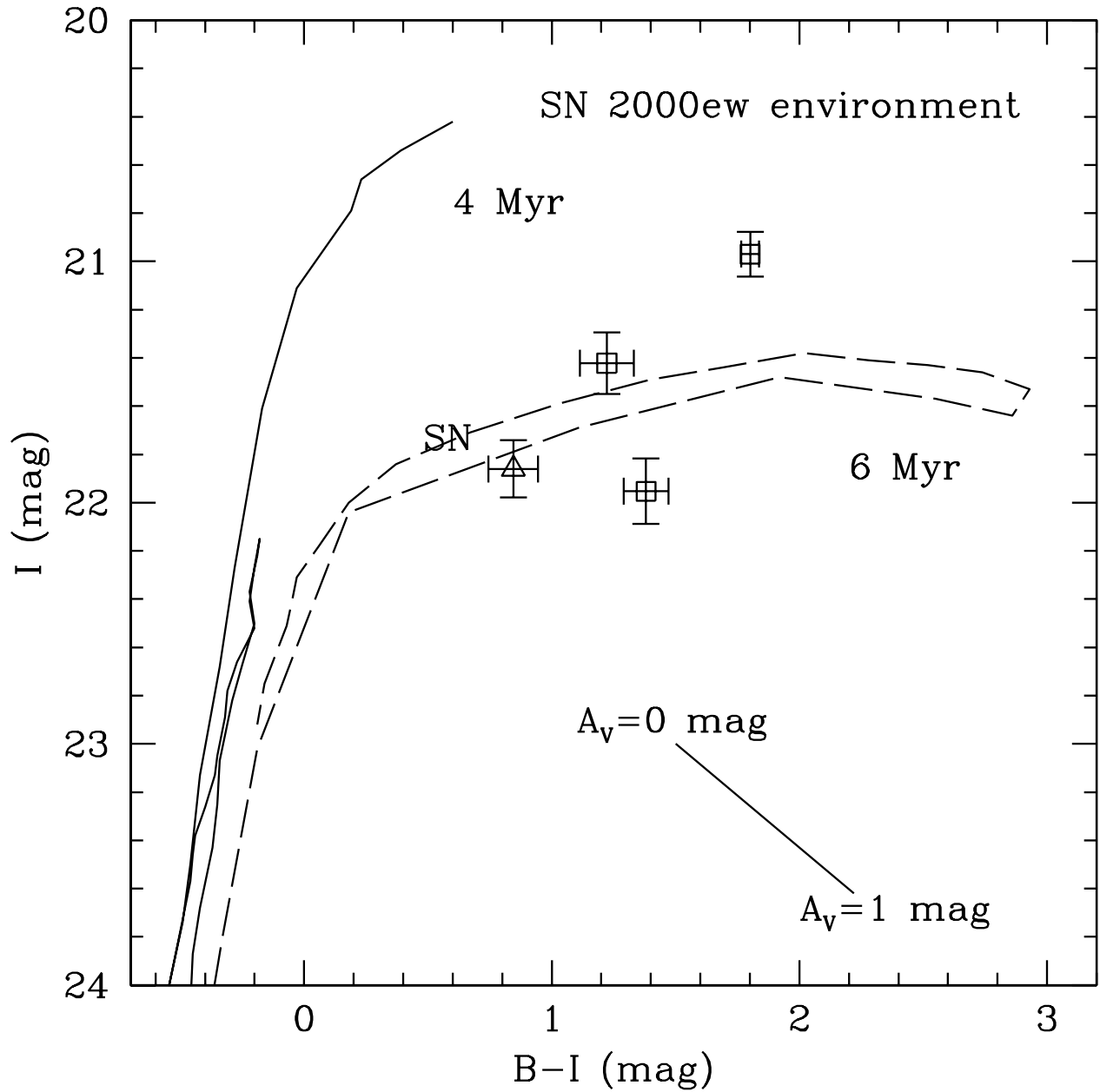}
\caption{The ($B-I$, $I$) color-magnitude diagram for the SN 2000ew
environment.  The detected objects are the SN itself ({\it open triangle}) and
the three objects ({\it open squares}) seen in immediate proximity to the SN
in Figure 14.  Also shown on the diagram are 4 and 6 Myr isochrones from
Bertelli et al.~(1994), adjusted for the assumed distance of 15 Mpc and
reddened, assuming the Galactic $A_V=0.15$ mag and the reddening
vector, following the Cardelli et al.~(1989) reddening law.}
\end{figure}

\clearpage

\begin{figure}
\figurenum{17}
\caption{The site of SN 2001B in IC 391 in an archival 70-s F555W image
from 1994 February 21.  The error circle has radius $0{\farcs}3$.  A star
with $M^0_V \approx -9.3$ mag, which could be the progenitor, is indicated with
tickmarks toward the northern edge of the circle.}
\end{figure}


\begin{figure}
\figurenum{18}
\caption{The site of SN 2001ai in NGC 5278 in an archival 260-s F814W image
from 2000 December 18.  The error circle has radius $0{\farcs}4$.  Objects
A and B along the edge of the circle are indicated with tickmarks.}
\end{figure}


\begin{figure}
\figurenum{19}
\caption{The site of SN 2001ai in NGC 5278 in an archival 1500-s F300W image
from 2000 December 18.  The error circle has radius $0{\farcs}4$. Objects
A--C along the edge of the circle are indicated with tickmarks.  Object C
is considered by HSTphot to be stellar and is not detected at F814W.}
\end{figure}


\begin{figure}
\figurenum{20}
\caption{The site of SN 2001ci in NGC 3079 in an archival 140-s F814W image
from 1999 March 4.  The error circle has radius $0{\farcs}3$.  Although the 
F606W image from 2001 January 21 is deeper (560 s), the F814W ($I$) image 
is less affected by dust extinction in the host galaxy (the SN may be 
extinguished by $A_V \approx 5$--6 mag) and would better reveal a reddened
progenitor.  The progenitor is not detected in any of the archive images.}
\end{figure}


\begin{figure}
\figurenum{21}
\caption{The site of SN 2001du in NGC 1365 in an archival 100-s F555W image
from 1995 January 15.  The error circle has radius $0{\farcs}9$.  Three
stars, A--C, within the circle are indicated with tickmarks.  Stars A and B are 
blue, while Star C, with $M^0_V\approx -6.9$ and $(V-I)^0\approx 1.0$ mag, is red, 
and is therefore a possible candidate for the progenitor of this SN II-P.}
\end{figure}


\begin{figure}
\figurenum{22}
\plotone{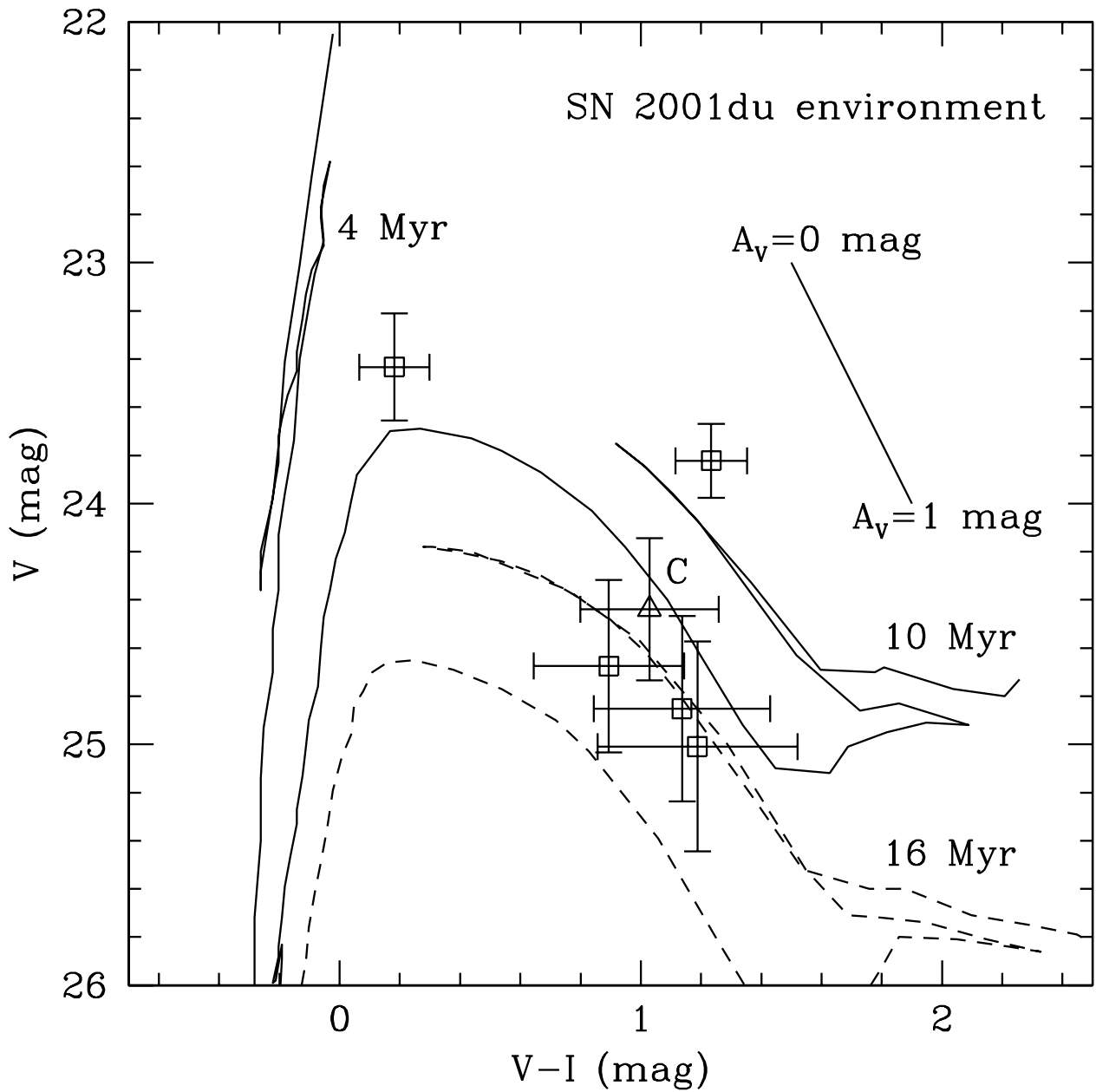}
\caption{The ($V-I$, $V$) color-magnitude diagram for the SN 2001du environment.
The progenitor candidate, Star C (see Figure 20), is represented with an {\it
open triangle}, and other stars with {\it open squares}.  (Stars A and B are not
detected at $I$.)  Also shown on the
diagram are 4, 10, and 16 Myr isochrones from Bertelli et al.~(1994) and the 
reddening vector, following the Cardelli et al.~(1989) reddening law.  The 
isochrones are adjusted for the assumed distance modulus $\mu=31.3$ mag
(Silbermann et al.~1999) and
reddened, assuming the Galactic $A_V=0.07$ mag.}
\end{figure}


\begin{figure}
\figurenum{23}
\caption{The site of SN 2001is in NGC 1961 in an archival 4000-s F547M image
from 2001 July 14.  The error circle has radius $0{\farcs}4$.  Two candidates
for the progenitor of this SN Ib, objects A and B, with $M^0_V\approx -8.3$ 
and $-8.7$ mag, respectively, are indicated with tickmarks.}
\end{figure}


\begin{thebibliography}{}
\bibitem[Aldering et al.~1994]{ald94} Aldering, G., Humphreys, R.~M., \&
Richmond, M. W. 1994, \aj, 107, 662

\bibitem[Armstrong 1999]{arm99} Armstrong, M. 1999, IAU Circ. 7306

\bibitem[Ayani et al.~1999]{aya99} Ayani, K., Furusho, R., Kawakita, H.,
Fujii, M., \& Yamaoka, H. 1999, IAU Circ. 7244

\bibitem[Barth et al.~1996]{bar96} Barth, A.~J., Van Dyk, S.~D., Filippenko,
A.~V., Leibundgut, B., \& Richmond, M.~W. 1996, AJ, 111, 2047

\bibitem[Benetti et al.~2001]{ben01} Benetti, S., Altavilla, G., Pastorello,
A., Turatto, M., Desidera, S., Giro, E., \& Cappellaro, E. 2001, IAU Circ. 7787

\bibitem[Bertelli et al.~1994]{ber94} Bertelli, G., Bressan, A., Chiosi,
C., Fagotto, F., \& Nasi, E. 1994, \aaps, 106, 275

\bibitem[Bertola 1964]{ber64} Bertola, F. 1964, Ann. d'Ap., 27, 319

\bibitem[Bessell 1990]{bes90} Bessell, M. S. 1990, \pasp, 102, 1181.

\bibitem[Beutler et al. 2002]{beu02} Beutler, B., Li, W. D., Filippenko, A. V.,
  Treffers, R. R., \& Schwartz, M. 2002, IAU Circ. 7906

\bibitem[Branch \& Nomoto 1986]{bra86} Branch, D., \& Nomoto, K. 1986, \aap,
164, L13

\bibitem[Branch, Nomoto, \& Filippenko 1991]{bra91} Branch, D., Nomoto, K.,
\& Filippenko, A.~V. 1991, Comm.~Ap., XV, 221

\bibitem[Branch et al.~2002]{bra02} Branch, D., et al. 2002, \apj, 566, 1005

\bibitem[Cao \& Gu 1999]{cao99} Cao, L., \& Gu, Q. S. 1999, IAU Circ. 7124

\bibitem[Cappellaro et al.~2000]{cap00} Cappellaro, E., Turatto, M., Pizzella,
A., Corsini, E.~M., Moro, D., \& Galletta, G. 2000, IAU Circ. 7352

\bibitem[Cardelli, Clayton, \& Mathis 1989]{car89} Cardelli, J. A.,
Clayton, G. C., \& Mathis, J. S. 1989, \apj, 345, 245

\bibitem[Chornock \& Filippenko 2001]{cho01} Chornock, R., \& Filippenko, A.~V.
2001, IAU Circ. 7577

\bibitem[Cohen, Darling, \& Porter 1995]{coh95} Cohen, J. G., Darling, J., \&
Porter, A. 1995, AJ, 110, 308

\bibitem[Deng et al.~2000]{den00} Deng, J. S., Qiu, Y. L., Hu, J. Y., Hatano,
K., \& Branch, D. 2000, \apj, 540, 452

\bibitem[Dolphin 2000a]{dol00a} Dolphin, A.~E. 2000a, \pasp, 112, 1383

\bibitem[Dolphin 2000b]{dol00b} Dolphin, A.~E. 2000b, \pasp, 112, 1397

\bibitem[Elmegreen et al.~2001]{elm01} Elmegreen, D.~M., Kaufman, M.,
Elmegreen, B.~G., Brinks, E., Struck, C., Klaric, M., \& Thomasson, M. 2001,
AJ, 121, 182

\bibitem[Evans 2001]{eva01} Evans, R. 2001, IAU Circ. 7690

\bibitem[Fesen \& Becker 1991]{fes91} Fesen, R.~A., \& Becker, R.~H. 1991, 
\apj, 371, 621

\bibitem[Fesen, Becker, \& Blair 1987]{fes87} Fesen, R.~A., 
Becker, R.~H., \& Blair, W.~P. 1987, \apj, 313, 378

\bibitem[Filippenko 1997]{fil97} Filippenko, A.~V. 1997, \araa, 35, 309

\bibitem[Filippenko \& Chornock 2000]{fil00a} Filippenko, A. V., \& Chornock, R.
2000, IAU Circ. 7511

\bibitem[Filippenko \& Chornock 2001]{fil01a} Filippenko, A. V., \& Chornock, R.
2001, IAU Circ. 7638

\bibitem[Filippenko, Chornock, \& Modjaz 2000]{fil00b} Filippenko, A. V.,
Chornock, R., \& Modjaz, M. 2000, IAU Circ. 7547

\bibitem[Filippenko, Leonard, \& Riess 1998]{fil98} Filippenko, A. V.,
Leonard, D. C., \&  Riess, A. G. 1998, IAU Circ. 6850

\bibitem[Filippenko \& Matheson 1993]{fil93a} Filippenko, A. V., \& Matheson, T.
1993, IAU Circ. 5842

\bibitem[Filippenko, Matheson, \& Ho 1993]{fil93b} Filippenko, A.~V., Matheson, 
T.,
\& Ho, L.~C. 1993, \apj, 415, L103

\bibitem[Filippenko, Stern, \& Reuland 1999]{fil99} Filippenko, A. V., Stern,
D., \& Reuland, M. 1999, IAU Circ. 7143

\bibitem[Filippenko et al.~1995]{fil95} Filippenko, A. V., Barth, A. J.,
Bower, G. C., Ho, L. C., Stringfellow, G. S., Goodrich, R. W., \&
Porter, A. C. 1995, AJ, 110, 2261 (Erratum: 1996, 112, 806)

\bibitem[Filippenko et al.~2001]{fil01b} Filippenko, A.~V., Li, W. D.,
Treffers, R. R., \& Modjaz, M. 2001, in
Small Telescope Astronomy on Global Scales (ASP Conf.~Ser.~246),
ed. B.~Paczy\'nski, W.-P. Chen, \& C. Lemme (San Francisco: ASP), 121

\bibitem[Foulkes et al. 2000]{fou00} Foulkes, S., Hurst, G., Migliardi, M.,
Villi, M., \& Li, W. D. 2000, IAU Circ. 7348

\bibitem[Friedman \& Li 1999]{fri99} Friedman, A., \& Li, W. 1999, IAU Circ. 7154

\bibitem[Garcia-Segura, Langer, \& Mac Low 1996]{gar96} Garc{\'\i}a-Segura, G., 
Langer, N., \& Mac Low, M.-M. 1996, \aap, 316, 133 

\bibitem[Garnavich et al.~1999a]{gar99a} Garnavich, P., Jha, S., Challis, P.,
Kirshner, R., \& Calkins, M. 1999a, IAU Circ. 7143

\bibitem[Garnavich et al.~1999b]{gar99b} Garnavich, P., Jha, S., Kirshner, R.,
Challis, P., \& Berlind, P. 1999b, IAU Circ. 7306

\bibitem[Garradd 2000]{gar00} Garradd, G.~J. 2000, IAU Circ. 7530

\bibitem[Gilmozzi et al.~1987]{gil87} Gilmozzi, R., et al.~1987, Nature, 328,
318

\bibitem[Goodrich et al.~1989]{goo89} Goodrich, R. W., Stringfellow, G. S.,
Penrod, G. D., \& Filippenko, A. V. 1989, \apj, 342, 908

\bibitem[Graham et al.~1997]{gra97} Graham, J.~A., et al.~1997, \apj, 477, 535

\bibitem[Hamuy \& Pinto 2002]{ham02}Hamuy, M., \& Pinto, P. A. 2002, \apjl,
566, L63

\bibitem[Ho, Filippenko, \& Sargent 1997]{lho97} Ho, L. C., Filippenko,
A. V., \& Sargent, W. L. W. 1997, \apjs, 112, 315

\bibitem[Humphreys \& Davidson 1979]{hum79} Humphreys, R. M., \& Davidson,
K. 1979, \apj, 232, 409

\bibitem[Hurst 1999]{hur99} Hurst, G. M. 1999, IAU Circ. 7306

\bibitem[Iwamoto et al.~1994]{iwa94} Iwamoto, K., Nomoto, K., H\"oflich, P.,
Yamaoka, H., Kumagai, S., \& Shigeyama, T. 1994, \apjl, 437, L115

\bibitem[Jacques 2001]{jac01} Jacques, C. 2001, IAU Circ. 7962

\bibitem[Jha et al.~1999a]{jha99a} Jha, S., Garnavich, P., Challis, P.,
Kirshner, R., \& Berlind, P. 1999a, IAU Circ. 7149

\bibitem[Jha et al.~1999c]{jha99c} Jha, S., Garnavich, P., Challis, P.,
Kirshner, R., \& Berlind, P. 1999c, IAU Circ. 7269

\bibitem[Jha et al.~1999b]{jha99b} Jha, S., Garnavich, P., Challis, P.,
Kirshner, R., Berlind, P., \& Howell, E. 1999b, IAU Circ. 7162

\bibitem[Jha et al.~2000]{jha00} Jha, S., Challis, P., Kirshner, R., \&
Berlind, P. 2000, IAU Circ. 7352

\bibitem[King 1999]{kin99} King, J. Y. 1999, IAU Circ. 7141

\bibitem[Klemola 1986]{kle86} Klemola, A.~R. 1986, \pasp, 98, 464

\bibitem[Krist 1995]{kri95} Krist, J. 1995, in Calibrating Hubble Space
Telescope: Post Servicing Mission (Baltimore: STScI), 311

\bibitem[Leonard et al.~2002a]{leo02a} Leonard, D.~C., et al.~2002a, \pasp,
  114, 35

\bibitem[Leonard et al.~2002b]{leo02b} Leonard, D.~C., et al.~2002b, \aj,
  114, 35

\bibitem[Li et al.~1998]{liw98} Li, W. D., Modjaz, M., Treffers, R. R., \&
Filippenko, A. V. 1998, IAU Circ. 6850

\bibitem[Li 1999]{liw99a} Li, W.~D. 1999, IAU Circ. 7145

\bibitem[Li et al.~1999]{liw99b} Li, W. D., Modjaz, M., King, J. Y.,
Papenkova, M., Johnson, R. A., Friedman, A., Treffers, R. R., \&
Filippenko, A. V. 1999, IAU Circ. 7126

\bibitem[Li et al.~2001]{liw01} Li, W.~D., et al. 2001, \pasp, 113, 1178

\bibitem[Li et al.~2002]{liw02} Li, W.~D., Filippenko, A.~V.,
Van Dyk, S.~D., Hu, J., Qiu, Y., Modjaz, M., \& Leonard, D.~C. 2002,
\pasp, 114, 403

\bibitem[Matheson et al.~2001a]{mat01a} Matheson, T., Filippenko, A. V.,
Li, W. D., Leonard, D. C., \& Shields, J. C. 2001a, AJ, 121, 1648

\bibitem[Matheson et al. 2001c]{mat01c} Matheson, T., Jha, S., Challis, P., 
Kirshner, R., \& Berlind, P. 2001c, IAU Circ. 7605

\bibitem[Matheson et al.~2001b]{mat01b} Matheson, T., Jha, S., Challis, P.,
Kirshner, R., \& Calkins, M. 2001b, IAU Circ. 7563

\bibitem[Modjaz \& Li 1999]{mod99} Modjaz, M., \& Li, W.~D. 1999, IAU Circ. 7268

\bibitem[Modjaz, Li, \& Schwartz 2001a]{mod01a} Modjaz, M., Li, W.~D., \&  
Schwartz, M. 2001a, IAU Circ. 7605

\bibitem[Modjaz et al.~2001b]{mod01b} Modjaz, M., Li, W. D., Filippenko, A.~V.,
King, J.~Y., Leonard, D.~C., Matheson, T., Treffers, R.~R., \& Riess, A.~G. 
2001b,
\pasp, 113, 308

\bibitem[Nomoto, Filippenko, \& Shigeyama 1990]{nom90} Nomoto, K., Filippenko,
A.~V., \& Shigeyama, T. 1990, \aap, 240, L1

\bibitem[Nomoto et al.~1994]{nom94} Nomoto, K., Yamaoka, H., Pols, O.~R.,
van den Heuvel, E.~P.~J., Iwamoto, K., Kumagai, S., \& Shigeyama, T. 1994,
Nature, 371, 227

\bibitem[Nomoto et al.~1996]{nom96} Nomoto, K., et al.~1996, in Compact Stars
in Binaries (IAU Symp.~165), ed.~J. van Paradijs, E. P. J. van den Heuvel, \&
E. Kuulkers (Dordrecht: Kluwer), 119

\bibitem[Pastorello et al. 1999]{pas99} Pastorello, A., Turatto, M., Rizzi, L.,
  Cappellaro, E., Benetti, S., \& Patat, F. 1999, IAU Circ. 7245

\bibitem[Patat et al.~1999]{pat99} Patat, F., Benetti, S., Cappellaro, E.,
Rizzi, L., \& Turatto, M. 1999, IAU Circ. 7183

\bibitem[Podsiadlowski et al.~1993]{pod93} Podsiadlowski, Ph., Hsu, J. J. L.,
Joss, P. C., \& Ross, R. R. 1993, Nature, 364, 509

\bibitem[Podsiadlowski et al.~1992]{pod92} Podsiadlowski, Ph., Joss, P. C.,
\& Hsu, J. J. L. 1992, \apj, 391, 246

\bibitem[Pogge \& Martini 2002]{pog02} Pogge, R.~W., \& Martini, P. 2002,
\apj, 569, 624

\bibitem[Puckett \& Dowdle 2000]{puc00a} Puckett, T., \& Dowdle, G. 2000,
IAU Circ. 7507

\bibitem[Puckett \& Langoussis 2000]{puc00b} Puckett, T., \& Langoussis, A.
2000, IAU Circ. 7530

\bibitem[Qiu et al.~2001]{qiu01} Qiu, Y. L., Hu, J. Y., Papenkova, M., \&
Schwartz, M. 2001, IAU Circ. 7782

\bibitem[Qiu et al.~1999]{qiu99} Qiu, Y. L., Qiao, Q. Y., Hu, J. Y., Zhou, X.,
\& Zheng, Z. 1999, IAU Circ. 7241

\bibitem[Ryder et al.~1993]{ryd93} Ryder, S., Staveley-Smith, L., Dopita, M.,
Petre, R., Colbert, E., Malin, D., \& Schlegel, E. M. 1993, \apj, 416, 167

\bibitem[Saha et al.~2001]{sah01} Saha, A., Sandage, A., Tammann, G.~A., 
Dolphin, A.~E., Christensen, J., Panagia, N., \& Macchetto, F.~D. 2001, 
\apj, 562, 314

\bibitem[Schlegel 1996]{sch96} Schlegel, E.~M. 1996, \apj, 111, 1660

\bibitem[Schlegel, Finkbeiner, \& Davis 1998]{sch98} Schlegel, D. J.,
Finkbeiner, D. P., \& Davis, M. 1998, \apj, 500, 525

\bibitem[Schwartz et al. 2000]{sch00} Schwartz, M., Li, W. D., Filippenko,
  A. V., Modjaz, M., \& Treffers, R. R. 2000, IAU Circ. 7514

\bibitem[Silbermann et al.~1999]{sil99} Silbermann, N.~A., et al. 1999,
\apj, 515, 1

\bibitem[Sonneborn, Altner, \& Kirshner 1987]{son87} Sonneborn, G., Altner, B.,
\& Kirshner, R.~P. 1987, \apjl, 323, L35

\bibitem[Stetson 1987]{ste87} Stetson, P. B. 1987, \pasp, 99, 191

\bibitem[Stetson 1992]{ste92} Stetson, P. B. 1992, in ADASS (ASP Conf.~Ser.~25),
ed.~D.M. Worrall, C.~Bimesderfer, \& J.~Barnes (San Francisco: ASP), 297

\bibitem[Swift, Li, \& Filippenko 2001]{swi01} Swift, B., Li, W.~D., \&
Filippenko, A. V. 2001, IAU Circ. 7618

\bibitem[Turatto et al.~1999]{tur99} Turatto, M., Rizzi, L., Salvo, M.,
Cappellaro, E., Benetti, S., \& Patat, F. 1999, IAU Circ. 7244

\bibitem[Uomoto 1986]{uom86} Uomoto, A. 1986, \apj, 310, L35

\bibitem[van den Heuvel 1994]{vdh94} van den Heuvel, E.~P.~J. 1994, in 
Interacting Binaries, ed.~H. Nussbaumer \& A. Orr (Berlin: Springer-Verlag),
263

\bibitem[van der Hucht 2001]{vdh01} van der Hucht, K.~A. 2001,
New Astronomy Reviews, 45, 135

\bibitem[Van Dyk 1992]{van92} Van Dyk, S. D. 1992, AJ, 103, 1788

\bibitem[Van Dyk, Hamuy, \& Filippenko 1996]{van96} Van Dyk, S. D., Hamuy,
M., \& Filippenko, A. V. 1996, AJ, 111, 2017

\bibitem[Van Dyk et al.~1999a]{van99a} Van Dyk, S.~D., Peng, C.~Y., Barth, 
A.~J.,
Filippenko, A.~V., Chevalier, R.~A., Fesen, R.~A., Fransson, C., Kirshner, 
R.~P.,
\& Leibundgut, B. 1999a, \pasp, 111, 313

\bibitem[Van Dyk et al.~1999b]{van99b} Van Dyk, S. D., Peng, C. Y., Barth,
A. J., \& Filippenko, A. V. 1999b, AJ, 118, 2331

\bibitem[Van Dyk et al.~2000]{van00} Van Dyk, S.~D., Peng, C.~Y., King, J.~Y.,
Filippenko, A.~V., Treffers, R.~R., Li, W. D., \& Richmond, M.~W. 2000, PASP, 112,
1532

\bibitem[Van Dyk et al.~2001]{van01} Van Dyk, S. D., Li, W.~D., Filippenko,
A.~V., \& Bock, G. 2001, IAU Circ. 7705

\bibitem[Van Dyk, Filippenko, \& Li 2002a]{van02a} Van Dyk, S. D., Filippenko,
A. V., \& Li, W.~D. 2002a, \pasp, 114, 701

\bibitem[Van Dyk et al.~2002b]{van02b} Van Dyk, S. D., Garnavich, P.~M.,
Filippenko, A. V., H\"oflich, P., Kirshner, R.~P., Kurucz, R.~L., \& Challis
P. 2002b, \pasp, in press

\bibitem[Wang et al.~2001]{wan01} Wang, L., Baade, D., Fransson, C.,
H\"oflich, P., Lundqvist, P., \& Wheeler, J. C. 2001, IAU Circ. 7704

\bibitem[Wei et al.~1999]{wei99} Wei, J. Y., Cao, L., Qiu, Y. L., Qiao, Q. Y.,
\& Hu, J. Y. 1999, IAU Circ. 7124

\bibitem[Woosley 1988]{woo88} Woosley, S.~E. 1988, \apj, 330, 218

\bibitem[Woosley \& Weaver 1986]{woo86} Woosley, S.~E., \& Weaver, T.~A. 1986,
\araa, 24, 205

\bibitem[Woosley \& Weaver 1995]{woo95} Woosley, S.~E., \& Weaver, T.~A. 1995,
\apjs, 101, 181

\bibitem[Woosley, Langer, \& Weaver 1993]{woo93} Woosley, S.~E., Langer, N.,
\& Weaver, T.~A. 1993, \apj, 411, 823

\bibitem[Xu \& Qiu 2001]{xud01} Xu, D. W., \& Qiu, Y. L. 2001, IAU Circ. 7555

\bibitem[Yasuda, Fukugita, \& Okamura 1997]{yas97} Yasuda, N., Fukugita,
M., \& Okamura, S. 1997, \apjs, 108, 417

\bibitem[Zampieri et al.~2002]{zam02} Zampieri, L., Pastorello, A., Turatto,
M., Cappellaro, E., Benetti, S., Altavilla, G., Mazzali, P., \& Hamuy, M.
2002, \mnras, in press

\bibitem[Zwicky 1964]{zwi64} Zwicky, F. 1964, \apj, 139, 514

\bibitem[Zwicky 1965]{zwi65} Zwicky, F. 1965, in Stars and Stellar Systems,
Vol. 8, Stellar Structure, ed. L. H. Aller \& D. B. McLaughlin (Chicago:
University of Chicago Press), p. 367.

\end{thebibliography}
\end{document}